\documentclass[conference]{IEEEtran}
\IEEEoverridecommandlockouts

\usepackage{cite}
\usepackage{amsmath,amssymb,amsfonts}
\usepackage{textcomp}
\usepackage{pifont}
\usepackage{graphicx}
\usepackage{colortbl}
\usepackage{xcolor}
\usepackage{algorithm}
\usepackage[noend]{algorithmic}
\usepackage{multirow}
\usepackage{multicol}
\usepackage{booktabs}
\usepackage{threeparttable}
\usepackage{comment}
\usepackage{bm}
\usepackage{enumerate}
\usepackage{diagbox}
\usepackage{flushend}
\usepackage{threeparttable}
\usepackage{url}
\usepackage{booktabs}
\usepackage{tablefootnote}
\usepackage{hyperref}
\usepackage{comment}

\usepackage{amsthm}
\theoremstyle{plain}

\newtheorem{definition}{Definition}

\usepackage{subfigure}
\usepackage{setspace}
\usepackage{epstopdf}
\usepackage{hyperref}
\newtheorem{remark}{Remark}

\renewcommand{\algorithmiccomment}[1]{\bgroup\hfill$\triangleright$~#1\egroup}

\newcommand{\colorR}[1]{\textcolor{red}{#1}}

\def\BibTeX{{\rm B\kern-.05em{\sc i\kern-.025em b}\kern-.08em
    T\kern-.1667em\lower.7ex\hbox{E}\kern-.125emX}}
\begin{document}

\title{PGB: Benchmarking Differentially Private Synthetic Graph Generation Algorithms}

\makeatletter
\newcommand{\linebreakand}{%
  \end{@IEEEauthorhalign}
  \hfill\mbox{}\par
  \mbox{}\hfill\begin{@IEEEauthorhalign}
}
\makeatother

\author{
Shang Liu$^{1,3}$,
Hao Du$^2$,
Yang Cao$^3$,
Bo Yan$^{4,3}$,
Jinfei Liu$^5$,
Masatoshi Yoshikawa$^{6}$
\vspace{5pt} \\

$^{1}$School of Computer Science and Technology, China University of Mining and Technology;\\
Mine Digitization Engineering Research Center of the Ministry of Education, Xuzhou, China, \\
$^2$Hokkaido University,
$^3$Institute of Science Tokyo, \\
$^4$Beijing University of Posts and Telecommunications,
$^5$Zhejiang University,
$^6$Osaka Seikei University 
\vspace{5pt} \\

shang@cumt.edu.cn,
hao.du.y4@elms.hokudai.ac.jp,
cao@c.titech.ac.jp,\\
boyan@bupt.edu.cn,
jinfeiliu@zju.edu.cn,
yoshikawa-mas@osaka-seikei.ac.jp
}




\maketitle

\begin{abstract}
Differentially private graph analysis is a powerful tool for deriving insights from diverse graph data while protecting individual information.
Designing private analytic algorithms for different graph queries often requires starting from scratch. 
In contrast, differentially private synthetic graph generation offers a general paradigm that supports one-time generation for multiple queries. 
Although various differentially private graph generation algorithms have been proposed, comparing them effectively remains challenging due to various factors, including differing privacy definitions, diverse graph datasets, varied privacy requirements, and multiple utility metrics.

To this end, we propose $\mathsf{PGB}$ (\underline{P}rivate \underline{G}raph \underline{B}enchmark), a comprehensive benchmark designed to enable researchers to compare differentially private graph generation algorithms fairly.
We begin by identifying four essential elements of existing works as a 4-tuple: mechanisms, graph datasets, privacy requirements, and utility metrics. 
We discuss principles regarding these elements to ensure the comprehensiveness of a benchmark.
Next, we present a benchmark instantiation that adheres to all principles, establishing a new method to evaluate existing and newly proposed graph generation algorithms. 
Through extensive theoretical and empirical analysis, we gain valuable insights into the strengths and weaknesses of prior algorithms. 
Our results indicate that there is no universal solution for all possible cases. 
Finally, we provide guidelines to help researchers select appropriate mechanisms for various scenarios.

\end{abstract}


\section{Introduction}
\label{sec:introduction}
Graph analysis serves as an effective method for deriving insights from diverse graph datasets, including social networks, traffic networks, and epidemiological networks. 
For instance, the degree distribution \cite{day2016publishing,liu2022crypto,hay2009accurate}, which counts the connections per node, illuminates the connectivity within social graphs. 
Additionally, subgraph counting
\cite{liu2024cargo,imola2021locally,liu2022collecting}, such as triangles or stars, aids in assessing central properties like the clustering coefficient \cite{newman2009random}, reflecting the probability that two connections of an individual are mutually linked. 
However, publicly sharing these graph statistics risks disclosing personal details \cite{RHMS_2}, as graph analytics are often conducted over sensitive information.

Differential privacy (DP) \cite{dwork2014algorithmic,li2016differential} has become the de-facto standard for privacy preservation, providing individual privacy against adversaries with arbitrary background knowledge.
Unlike previous privacy definitions (e.g., $k$-anonymity, $l$-diversity, $t$-closeness), DP ensures that modifications of a single node or edge have a minimal impact on the output results.
Many differentially private graph analytic algorithms have been designed for various graph queries, such as degree distribution~\cite{liu2022crypto,day2016publishing,hay2009accurate}, subgraph counts~\cite{imola2021locally,liu2024cargo,liu2022collecting}, and community detection~\cite{nguyen2016detecting,mohamed2022differentially,fu2024community}.
Unfortunately, these solutions are usually tailored to specific graph queries.
For different queries, differentially private graph algorithms must be designed from scratch.
One solution is to privately generate a synthetic graph that maintains semantic similarity to the original graph while satisfying DP.
This paradigm is superior to tailored algorithms as it enables one-time generation for multiple queries.

Despite a rich set of differentially private synthetic graph generation algorithms \cite{wang2013preserving,nguyen2015differentially,chen2014correlated,mir2012differentially,xiao2014differentially,yuan2023privgraph,chen2019publishing,jorgensen2016publishing,zhang2020community,jian2021publishing,qin2017generating,ju2019generating,ye2020lf,wei2020asgldp,hou2023block,liu2020local} having been proposed, there is no generally acknowledged and unified procedure to perform empirical studies on them.
Concretely, it is challenging to compare them effectively due to the following factors:

\begin{itemize}
    \item 
    Algorithms use different privacy definitions to protect individual information in a graph, such as edge differential privacy \cite{wang2013preserving,nguyen2015differentially,chen2014correlated,mir2012differentially,xiao2014differentially,yuan2023privgraph,chen2019publishing,jorgensen2016publishing} and node differential privacy \cite{zhang2020community,jian2021publishing}.
    It is unfair to compare algorithms based on different privacy definitions.

    \item 
    Few algorithms in our literature survey offer open-source support. 
    Correctly re-implementing differentially private graph generation algorithms can be challenging due to their intrinsic complexity.

    \item 
    Many algorithms in publications exhibit data-dependent errors. 
    Their utility depends on the choice of input graph characteristics, such as graph size, average clustering coefficient, and graph type.
    
    \item 
    Algorithms are often associated with the privacy parameter $\varepsilon$, achieving optimal utility under different privacy requirements. 
    For example, DP-2K \cite{wang2013preserving} exhibits lower error than DK-1K \cite{wang2013preserving} on one graph when $\varepsilon > 20$; however, the results reverse when $\varepsilon \leq 20$.

    \item 
    All algorithms in our literature review cover only a subset of graph queries. 
    Additionally, even when evaluating the same query, different algorithms employ different error metrics. 
    For instance, PrivHRG \cite{xiao2014differentially} uses normalized mutual information \cite{kvalseth1987entropy} to measure the utility of community detection, whereas LF-GDPR \cite{ye2020lf} uses the adjusted random index \cite{rand1971objective} and adjusted mutual information \cite{vinh2009information}.
    
\end{itemize}

In this paper, we aim to address the aforementioned challenges with a comprehensive benchmark, $\mathsf{PGB}$ (\underline{P}rivate \underline{G}raph \underline{B}enchmark). 
Our contributions are summarized as follows:

\textit{Benchmark Design Principles.}
Based on a comprehensive literature review, we identify four essential elements of existing studies as a 4-tuple (\textbf{M}, \textbf{G}, \textbf{P}, \textbf{U}): mechanisms, graph datasets, privacy requirements, and utility metrics. 
For each element, we discuss the limitations of existing works and propose requirements to ensure comparable results (see more details in Section \ref{sec:benchmark principle}).

\textit{Benchmark Instantiation.}
We introduce the benchmark $\mathsf{PGB}$ to evaluate the utility of differentially private graph generation algorithms while adhering to all design principles. 
Our benchmark is implemented, and the source code is publicly available\footnote{\label{footnote:code}PGB code: \url{https://github.com/dooohow/PGB}}.
We also implement a benchmark platform\footnote{PGB platform: \url{https://pgb-result.github.io/}}, so future works can be included and compared easily (details in Section \ref{sec:benchmark instantiation}).

\textit{Empirical Study and Findings.}
We have conducted the largest empirical evaluation of private graph generation algorithms so far.
Based on our benchmark, it has at least 43,200 single experiments comprising 6 selected algorithms, 8 graph datasets, 6 privacy budgets, and 15 queries.
Our findings suggest that while some generation algorithms are generally strong performers, there is no one-size-fits-all solution.
For the complete paper, please refer to the version available on arXiv\footnote{PGB paper: \url{https://arxiv.org/abs/2408.02928}}
(details in Section~\ref{sec:experiment}).

\section{Related Works}
\label{sec:related works}

\subsection{Private Graph Generation}
There are multiple existing studies focusing on differentially private graph generation algorithms~\cite{wang2013preserving,nguyen2015differentially,chen2014correlated,mir2012differentially,xiao2014differentially,yuan2023privgraph,chen2019publishing,jorgensen2016publishing,zhang2020community,jian2021publishing,qin2017generating,ju2019generating,ye2020lf,wei2020asgldp,hou2023block,liu2020local,gao2019phdp,eliavs2020differentially,brito2023global}.
For instance, Gao \textit{et al.} \cite{gao2019phdp} introduce persistent homology for publishing online social networks. 
However, their approach lacks protection for the distance matrix, which may compromise individual privacy. 
Marek \textit{et al.} \cite{eliavs2020differentially} and Felipe \textit{et al.} \cite{brito2023global} focus on releasing attributed graphs or weighted graphs under DP.
In our evaluation, we consider five state-of-the-art works: DP-dK \cite{wang2013preserving}, TmF \cite{nguyen2015differentially}, PrivSKG \cite{mir2012differentially}, PrivHRG \cite{xiao2014differentially}, and PrivGraph \cite{yuan2023privgraph}, as well as one baseline approach DGG~\cite{qin2017generating}.

\textbf{DP-dK.}
DP-dK first condenses the graph into the degree distribution of K-connected components (dk-series). 
It then adds Laplace noise to the learned parameters and generates synthetic graphs with the perturbed parameters using the dK-series model \cite{mahadevan2006systematic}. 
For the DP-2K model, noise is calibrated based on smooth sensitivity rather than global sensitivity, resulting in noise of a smaller magnitude. 
Despite these improvements, the privacy budget required remains unreasonably large (i.e., $\varepsilon \geq 100$).

\textbf{TmF.}
It first represents a graph as an adjacency matrix, then adds Laplace noise to each cell. 
Finally, TmF selects the top-$m$ noisy cells as the edges in the randomized adjacency matrix, where $m$ is the noisy number of edges. 
However, most of the true edges cannot be retained from the top-$m$ noisy cells, especially when $\varepsilon$ is small.

\textbf{PrivSKG.}
It uses the stochastic Kronecker graph model to represent a graph and then constructs a private estimator of the true parameters. 
This private estimator defines a probability distribution over the graph. 
Finally, PrivSKG generates a synthetic graph by sampling from this distribution. 
Nevertheless, PrivSKG cannot accurately capture the structural properties of the true graph, as the generation process is determined by a single parameter.

\textbf{PrivHRG.}
PrivHRG first leverages a statistical hierarchical random graph (HRG) model \cite{clauset2008hierarchical} to represent a graph, recording connection probabilities between any pair of nodes. 
It then privately samples a dendrogram via Markov-Chain Monte Carlo (MCMC) \cite{metropolis1953equation}. 
Finally, the synthetic graph is generated based on the noisy connection probabilities. 
However, partial information of the true graph can be lost during the construction of the HRG model.

\textbf{PrivGraph.}
It first generates a coarse node partition using a community detection algorithm and applies the Exponential mechanism to obtain the community partitions privately. 
Then, PrivGraph computes the degree sequences within communities and the number of edges between communities. 
Finally, it generates a synthetic graph based on the noisy degree sequences using the CL model \cite{aiello2000random}. 
Compared with prior works, PrivGraph preserves more structural information of a graph by exploiting community information.

\textbf{DGG.}
Node degree is fundamental information in a graph and has been used for private graph generation \cite{qin2017generating,ye2020lf}. 
We revise DGG \cite{qin2017generating} to satisfy Edge CDP as our benchmark baseline. 
Specifically, DGG first calculates the node degrees and then perturbs these degrees using the Laplace mechanism. 
Finally, it generates a synthetic graph using the BTER model~\cite{seshadhri2012community}. 
However, DGG fails to capture the graph structure beyond node degrees, thereby losing detailed information about the true graph.

\begin{remark} 
A limited number of studies \cite{hou2023wdp,yang2020secure} generate synthetic graphs under differential privacy using deep learning (DL) methods (e.g., GANs). 
We exclude these studies from our benchmark for the following reasons.
1) Their privacy goals differ from those of the algorithms in our benchmark. 
Most algorithms in our benchmark focus solely on preserving graph structure information, whereas prior DL-based work~\cite{hou2023wdp,yang2020secure} consider both graph structure and node features. 
Incorporating node features into the training process requires additional privacy budget allocation.
2) The types of graph queries also differ. 
Synthetic graphs generated by DL-based methods are evaluated primarily through deep learning tasks, such as link prediction, which differ from the statistical queries in our benchmark.
\end{remark}

\subsection{DP Benchmarks}
DP benchmarks on data analysis have recently received much attention from researchers, encompassing both \textit{graph data} and \textit{tabular data}.
Ning \textit{et al.} \cite{ning2021benchmarking} implement and benchmark various graph queries (i.e., degree distribution and subgraph counting) by examining the trade-offs between privacy, accuracy, and performance.
These implementations of private graph algorithms have been integrated into DPGraph \cite{xia2021dpgraph}.
DPGraph is a benchmark platform for differentially private graph analysis.
This platform helps researchers understand the trade-offs between privacy, accuracy, and performance of existing private graph analysis algorithms, primarily focusing on degree distribution and subgraph counting.
These benchmarks motivate us to design a comprehensive benchmark for differentially private synthetic graph generation algorithms.

In addition, there are many benchmarks on differentially private tabular data analysis.
DPBench \cite{hay2016principled} is a principled framework for evaluating differential privacy algorithms, such as 1- and 2-dimensional range queries.
DPComp \cite{hay2016exploring} is a publicly accessible web-based system to support the principled evaluation of private data analysis and to encourage the dissemination of related code and data.
Tao \textit{et al.} \cite{tao2021benchmarking} propose a systematic benchmark on differentially private synthetic tabular data generation algorithms, including GAN-based, marginal-based, and workload-based methods.
Basu \textit{et al.} \cite{basu2021benchmarking} design a benchmark on the utility of central and federated training of BERT-based models using depression and sexual harassment-related Tweets.
Schäler \textit{et al.} \cite{schaler2023benchmarking} introduce a comparable benchmark that meets all design requirements. They conduct the largest empirical study on $w$-event differential privacy mechanisms.
Rosenblatt \textit{et al.}~\cite{rosenblatt2024epistemic} propose an evaluation methodology for DP synthesizers based on reproducibility.
Gonzalo \textit{et al.} \cite{garrido2021get} conducted a comprehensive comparison and evaluation of five mainstream open-source DP libraries.
Dmitry \textit{et al.}\cite{prokhorenkov2023towards} reviewed recent studies on existing attack types, as well as methods and metrics used to assess privacy risks. 
But, these benchmarks cannot be directly used to evaluate graph data due to the unique characteristics of graphs, such as privacy definitions, representations, and utility metrics.

\section{Preliminary}
\label{sec:preliminary}

\subsection{Differential Privacy}
Differential privacy (DP)  \cite{dwork2014algorithmic,li2016differential} has become a de-facto standard for preserving individual privacy.
In the context of graphs, which are composed of nodes and edges, DP can be defined in two ways: \emph{edge differential privacy} (Edge DP) and \emph{node differential privacy} (Node DP) \cite{hay2009accurate}.
Edge DP ensures that the output of a randomized mechanism does not reveal whether any specific friendship information (i.e., edge) exists in a graph. In contrast, Node DP conceals the existence of a particular user (i.e., node) along with all her adjacent edges. Node DP provides a stronger privacy guarantee because it protects both node and edge information. However, this stronger privacy comes at the cost of utility.
Based on different assumptions, we have the following definitions: Node CDP, Edge CDP, and Edge LDP.

\begin{definition}[Differential Privacy \cite{dwork2014algorithmic}]
	\label{def:DP}
Let $\varepsilon > 0$ be the privacy budget.
A randomized algorithm $\mathcal{M}$ with domain $\mathcal{X}$ satisfies $\varepsilon$-DP, if for any neighboring databases $D, D^\prime \in \mathcal{X}$ that differ in a single datum and any subset $S \subseteq Range(\mathcal{M})$, 
	\begin{center}
		$Pr[\mathcal{M}(D) \in S] \leq e^{\epsilon} Pr[\mathcal{M}(D^{\prime}) \in S]$
	\end{center}  
\end{definition}

\begin{figure*}[t]
	\centering  
	\includegraphics[width=0.7\linewidth]{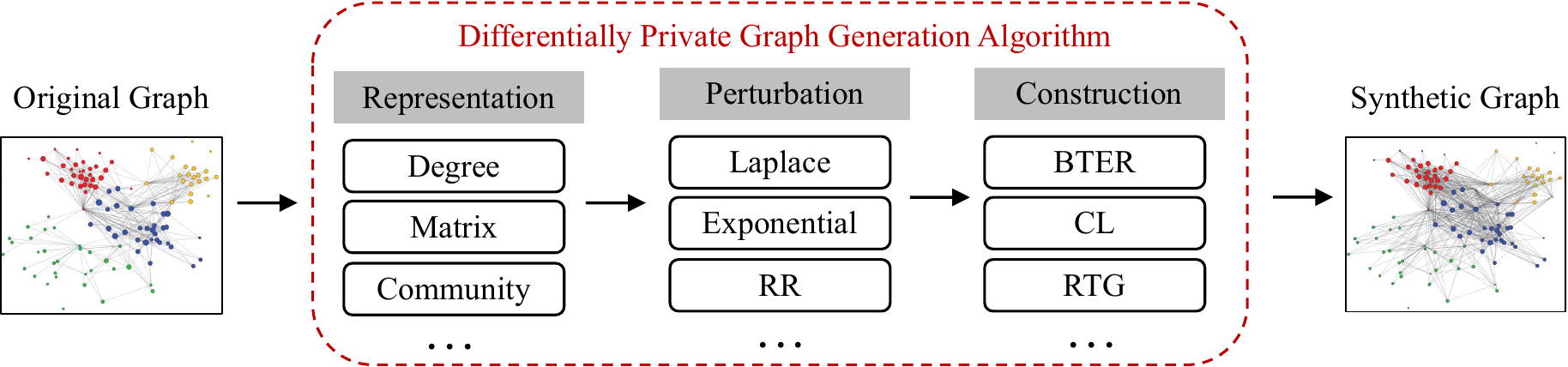}
	\caption{The common steps for differentially private graph generation algorithms: Representation, Perturbation, and Construction.}
	\label{fig:framework}  
\end{figure*}

\begin{definition}[Node CDP \cite{hay2009accurate}]
	\label{def:nodeDP}
Let $\varepsilon > 0$ be the privacy budget.
A randomized algorithm $\mathcal{M}$ with domain $\mathcal{G}$ satisfies $\varepsilon$-Node DP, if for any two neighboring graphs $G, G^\prime \in \mathcal{G}$ that differ in one node with all edges incident to it, and any subset $S \subseteq Range(\mathcal{M})$, 
	\begin{center}
	$Pr[\mathcal{M}(G) \in S] \leq e^{\epsilon} Pr[\mathcal{M}(G^\prime) \in S]$
	\end{center} 
\end{definition}

\begin{definition}[Edge CDP \cite{raskhodnikova2016differentially}]
	\label{def:edgeDP}
Let $\varepsilon > 0$ be the privacy budget.
A randomized algorithm $\mathcal{M}$ with domain $\mathcal{G}$ satisfies $\varepsilon$-Edge CDP, iff for any two neighboring graphs $G, G^\prime \in \mathcal{G}$ that differ in one edge and any subset $S \subseteq Range(\mathcal{M})$, 
	\begin{center}
	$Pr[\mathcal{M}(G) \in S] \leq e^{\epsilon} Pr[\mathcal{M}(G^\prime) \in S]$
	\end{center} 
\end{definition}

\begin{definition}[Edge LDP \cite{qin2017generating}]
	\label{def:edgeLDP}
Let $\varepsilon > 0$ be the privacy budget.
For any $i\in[n]$, let $\mathcal{M}_i$ be a randomized algorithm of user $v_i$. 
$\mathcal{M}_i$ satisfies $\varepsilon$-Edge LDP, iff for any two neighboring adjacent bit vectors $A_i$ and $A_i^\prime$ that differ in one edge and any subset $S \subseteq Range(\mathcal{M}_i)$, 
	\begin{center}
	$Pr[\mathcal{M}_i(A_i) \in S] \leq e^{\epsilon} Pr[\mathcal{M}_i(A_i^\prime) \in S]$
	\end{center} 
\end{definition}

\subsection{Graph Synthesis with DP}
We now introduce a common framework for differentially private graph generation, designed to encompass all mechanisms of our literature survey.
This framework enables us to compare mechanisms both theoretically and empirically.
As shown in Figure \ref{fig:framework}, the common framework of differentially private graph synthesis models consists of three main stages: representation, perturbation, and construction.

\textit{Representation.}
The first stage involves modeling the original graph and identifying a compact representation. 
Various representations, such as degree information \cite{wang2013preserving,qin2017generating,ye2020lf}, adjacency matrix \cite{nguyen2015differentially,chen2014correlated,mir2012differentially}, or community structure \cite{yuan2023privgraph,ju2019generating,chen2019publishing,liu2020local}, are used to capture the essential properties of the graph. 
It is worth noting that the compact representation effectively addresses the challenge of high-dimensional graph data by reducing the added noise required to guarantee DP.

\textit{Perturbation.}
In this stage, suitable noise is added to the compact representation to satisfy the differential privacy.
Common randomized mechanisms include the Laplace mechanism~\cite{dwork2006calibrating}, Exponential mechanism \cite{mcsherry2007mechanism}, Randomized Response (RR) \cite{warner1965randomized}, and so forth.
According to the post-processing property \cite{dwork2014algorithmic}, subsequent processes of graph synthesis do not compromise individual privacy.

\textit{Construction.}
The final stage involves constructing a synthetic graph from the perturbed representations. 
Some graph constructors such as Block Two-level Erdős-Rényi (BTER)~\cite{seshadhri2012community} and Chung-Lu (CL) \cite{aiello2000random} are employed to construct synthetic graphs while preserving the desired structural properties.
In fact, graph constructor models have been widely discussed in research communities \cite{bonifati2020graph}.
A vast majority of graph constructors are created for different requirements.
Publications in our literature survey use different constructors to generate graphs.
For instance, LDPGen \cite{qin2017generating} leverages the BTER model and PrivGraph \cite{yuan2023privgraph} uses the CL model. 

\begin{remark}
    We treat differentially private graph generation algorithms as black boxes, aiming to provide motivation for selecting them in various scenarios. 
    Thus, the choices made at each step (i.e., representation, perturbation, and construction) within the algorithms are beyond the scope of our benchmark.
\end{remark}

\section{Benchmark Design Principles}
\label{sec:benchmark principle}

In this section, we outline the fundamental design principles that underpin our benchmarking framework $\mathsf{PGB}$ (see Section~\ref{sec:benchmark instantiation}). 
These principles are critical for ensuring comprehensive, fair, and meaningful comparisons of differentially private graph synthesis algorithms.
Prior works often neglect these principles, leading to incomplete or biased evaluations.
To develop a robust benchmarking framework, we conducted a thorough literature review as listed in Table \ref{tab:comparisons of papers}, encompassing key publications from notable conferences or journals such as CCS, VLDB, SIGMOD, and TKDE. 
We identified essential elements of empirical studies as a 4-tuple (\textbf{M}, \textbf{G}, \textbf{P}, \textbf{U}):
\begin{itemize}
    \item \textbf{M}: A set of mechanisms being compared.
    \item \textbf{G}: A set of graph datasets.
    \item \textbf{P}: A set of privacy requirements.
    \item \textbf{U}: A set of utility metrics.
\end{itemize}

Next, we delve into these elements in detail and discuss the requirements necessary to ensure the comprehensiveness of designing a benchmark. 

\begin{table*}[t]
\small
	\caption{Comparisons of previous works.}
	\label{tab:comparisons of papers}
	\centering
 \resizebox{2.05\columnwidth}{!}{
		\begin{tabular}{|l|cccc|cccl|c|l|l|}
			\hline
     \multirow{2}{*}[-0.1em]{Algorithm} & \multicolumn{4}{c|}{Mechanism (\textbf{M})} & \multicolumn{4}{c|}{Graph (\textbf{G})} & Privacy (\textbf{P}) & \multicolumn{2}{c|}{Utility (\textbf{U})} \\
     \cline{2-12} 
       & P.D. & $\Delta$ & Attr.  & Code & $|\text{V}|$ ($10^x$) & $|\text{E}|$  ($10^x$) &  ACC & Type & $\varepsilon$ & \multicolumn{1}{c|}{Query}  & \multicolumn{1}{c|}{Metric} \\
			\hline
    DP-dK \cite{wang2013preserving} & E.C.  & S & \ding{55} &   \ding{55} & $2 \sim 3$ & $2 \sim 4$ & 0.25 $\sim$ 0.63 & $\text{T}_{1,3,6}$ & [0.2,2000] & $\text{Q}_{1\sim 4,7,8,11,13,15}$ & $\text{E}_{1}$ \\
    
    TmF \cite{nguyen2015differentially} & E.C.  & G & \ding{55} & \ding{55} & $2 \sim 6$ & $2 \sim 6$ & 0.25 $\sim$ 0.63 & $\text{T}_{1\sim 3,5}$ & (0, 50) & $\text{Q}_{4\sim 10}$ & $\text{E}_{1}$\\
    
    DER \cite{chen2014correlated} & E.C.  & G & \ding{55} & \ding{55} & $3$ & $3 \sim 5$ & 0.14 $\sim$ 0.61 & $\text{T}_{1,3,4}$ & (0.6,1) & $\text{Q}_{1,6,8}$ & $\text{E}_{2,3}$\\
    
    PrivSKG \cite{mir2012differentially} & E.C.  & S & \ding{55} & \ding{55} & $3 \sim 4$ & $4 \sim 5$ & 0.25 $\sim$ 0.61 & $\text{T}_{2,3,7}$ & 0.2 & $\text{Q}_{6,11,15}$ & -\\
     
    PrivHRG \cite{xiao2014differentially} & E.C.  &  G & \ding{55} &      \colorR{\ding{51}} & $3 \sim 5$ & $4 \sim 5$ & 0.14 $\sim$ 0.63 & $\text{T}_{1\sim 3}$ & 1 & $\text{Q}_{6,9,15}$ & $\text{E}_{7}$\\
    
    PrivGraph \cite{yuan2023privgraph} & E.C.   & G & \ding{55} & \colorR{\ding{51}} & $3 \sim 5$ & $4 \sim 5$ & 0.13 $\sim$ 0.61 & $\text{T}_{1\sim 3}$ & [0.5,3.5] & $\text{Q}_{6,7,10,12,15}$ & $\text{E}_{1,3,7,11}$ \\
    
    C-AGM \cite{chen2019publishing} & E.C.  & G & \colorR{\ding{51}} & \ding{55} & $3 \sim 4$ & $4 \sim 5$ & 0.13 $\sim$ 0.54 & $\text{T}_{1\sim 3}$ & [2, 9] & $\text{Q}_{2,3,6,10}$ & $\text{E}_{1,4,6}$\\
    
    TriCycLe \cite{jorgensen2016publishing} & E.C.  & S & \colorR{\ding{51}} & \ding{55} & $3 \sim 5$ & $4 \sim 6$ & 0.10 $\sim$ 0.18 & $\text{T}_{1\sim 3}$ & [0.01,ln3] & $\text{Q}_{2,3,6,10,11}$ & $\text{E}_{2,5}$ \\
    
    PrivCom \cite{zhang2020community} & N.C.   & G & \ding{55} & \ding{55} & $3$ & $4$ & 0.52 & $\text{T}_{1\sim 3}$ & [0.1, 20] & $\text{Q}_{12}$ & $\text{E}_{6}$ \\
    
    $\pi_v,\pi_e$ \cite{jian2021publishing} & N.C.  & G & \ding{55} & \ding{55} & $3 \sim 6$ & $4 \sim 7$ & 0.11 $\sim$ 0.61 & $\text{T}_{1,3,7}$ & [0.1, 20] & $\text{Q}_{1\sim 3,6,10,11}$ & $\text{E}_{2,4}$\\
    
    LDPGen \cite{qin2017generating} & E.L.   & G & \ding{55} & \ding{55} & $3 \sim 5$ & $4 \sim 5$ & 0.49 $\sim$ 0.61 & ${\text{T}_1}$ & (0, 7] & $\text{Q}_{10,12\sim 14}$ & $\text{E}_{1,9,10}$\\
    
    CGGen \cite{ju2019generating} & E.L.  & G  & \ding{55} & \ding{55} & $3 \sim 4$ & $4 \sim 5$ & 0.49 $\sim$ 0.61 & ${\text{T}_1}$ & (0, 7] & $\text{Q}_{10,13,14}$ & $\text{E}_{1,9,10}$\\
    
    LF-GDPR \cite{ye2020lf} & E.L.  & G  & \ding{55} & \colorR{\ding{51}} & $3 \sim 5$ & $4 \sim 7$ & 0.49 $\sim$ 0.63 & $\text{T}_{1,3}$ & [1, 8] & $\text{Q}_{10,12,13}$ & $\text{E}_{1,8,9,10}$\\
    
    AsgLDP \cite{wei2020asgldp} & E.L.  & G  & \colorR{\ding{51}} & \ding{55} & $3 \sim 5$ & $4 \sim 7$ & 0.49 $\sim$ 0.61 & ${\text{T}_1}$ & [0.1,9] & $\text{Q}_{6,10,13}$ & $\text{E}_{1,4}$\\
    
    Block-HRG \cite{hou2023block} & E.L.  & G & \ding{55} & \ding{55} & $3 \sim 4$ & $4 \sim 5$ & 0.49 $\sim$ 0.63 & $\text{T}_{1,3,6}$ & [1, 8] & $\text{Q}_{4,6,10,11,13}$ & $\text{E}_{1,9,10}$\\
    
    DP-LUSN \cite{liu2020local} & E.L.  & G & \ding{55} & \ding{55} & $2 \sim 3$ & $3$ & - & $\text{T}_{2,3}$ & [0.1, 1] & $\text{Q}_{2,10}$ & - \\
	 \hline
	\end{tabular}
 }
    \begin{tablenotes}
     \item[1] P.D.: Privacy Definition \quad E.C.: Edge CDP \quad N.C.: Node CDP \quad E.L.: Edge LDP \quad $\Delta$: Sensitivity \quad G: Global \quad S: Smooth 
     \item[2] $|\text{V}|$: Number of Nodes \quad $|\text{E}|$: Number of Edges \quad ACC: Average Clustering Coefficient \quad $\varepsilon$: Privacy Budget \quad \colorR{\ding{51}}: yes \quad \ding{55}: no
     \item[3] Table \ref{tab:graph type}, Table \ref{tab:graph query}, and Table \ref{tab:symnbols of graph queries} provide details for Type, Query, and Metric, respectively.
   \end{tablenotes}
\end{table*}

\subsection{Mechanisms \textbf{M}}
We consider 4 principles (\textbf{M$_1$}$\sim$\textbf{M$_4$}) that the mechanism \textbf{M} should satisfy to ensure fair comparisons. 
Additionally, we discuss the extent to which existing works adhere to these principles, as summarized in Table \ref{tab:comparisons of papers}.

\subsubsection{Privacy Definition (\textbf{M$_1$})}
A fair comparison of algorithms needs identical privacy definitions.
When DP is applied to graph analysis, we have two kinds of privacy definitions since a graph consists of nodes and edges: edge differential privacy and node differential privacy \cite{hay2009accurate}.
The former guarantees that a randomized mechanism does not disclose the addition or deletion of a specific edge belonging to an individual \cite{blocki2012johnson}, while the latter obscures the addition or deletion of a node and all its connected edges \cite{kasiviswanathan2013analyzing}.
Besides, there are two other privacy definitions based on different trust assumptions (users trust or untrust the server): central differential privacy (CDP) and local differential privacy (LDP).
In CDP\cite{dwork2008differential}, a trusted server collects all original data from each user to compute and perturb the query results.
In LDP, \cite{kasiviswanathan2011can}, each user randomizes the data to ensure local DP directly.
Therefore, we have four privacy definitions to protect individual information in graph analysis: edge CDP, edge LDP, node CDP, and node LDP.
Our literature study (refer to Table \ref{tab:comparisons of papers}) reveals that half of 16 publications satisfy edge CDP;
6 out of 16 publications satisfy edge LDP;
only 2 studies satisfy CDP and no publication satisfies node LDP for private graph synthesis.
It is worth noting that algorithms cannot be comparable since they use different privacy definitions .
For example, node CDP provides a stronger privacy guarantee than edge CDP but at the cost of utility.
Similarly, edge CDP provides a higher utility than edge LDP but relies on a trust server.

\subsubsection{Sensitivity (\textbf{M$_2$})}
Our literature review reveals that the majority of publications use \textit{global sensitivity} \cite{dwork2006calibrating} to determine the magnitude of the added noise. 
In contrast, only three publications (DP-dK \cite{wang2013preserving}, PrivSKG \cite{mir2012differentially}, TriCycLe \cite{jorgensen2016publishing}) utilize \textit{smooth sensitivity} \cite{nissim2007smooth}.
Global sensitivity considers \textit{any} two neighboring graphs, which can be pessimistic since it covers the largest difference of all cases. 
Alternatively, \textit{local sensitivity} \cite{nissim2007smooth} fixes one graph and considers all of its neighbors. 
However, local sensitivity can potentially leak sensitive information about the fixed graph.
To address this, smooth sensitivity employs a “smooth approximation” of local sensitivity to calibrate the noise, thereby satisfying differential privacy (DP). 
It is acceptable for different algorithms to use various sensitivity definitions to measure the added noise. 
However, the premise is that they should provide identical privacy definitions to ensure the compatibility of the benchmark.

\subsubsection{Consideration of Attributed Graph (\textbf{M$_3$})}
There are rich but sensitive node attributes and edge attributes in real-world graphs.
For example, in a disease transmission analysis, we need to collect reports on each person's health condition (i.e., age, gender, and trajectory) and details of the disease transmission (i.e., transmission time, transmission method, and infection probability).
Our literature review indicates that most studies have focused on purely structured graphs, 
only a few algorithms \cite{chen2019publishing, jorgensen2016publishing, wei2020asgldp} consider graphs with node attributes, and no studies focus on graphs with edge attributes.
Directly comparing algorithms for attributed and non-attributed graphs may be unfair, as a portion of the privacy budget must be allocated to protect attributes. 
One solution is to transform an attributed graph synthesis algorithm into a non-attributed one, allowing the entire privacy budget to be used for protecting structural information.

\subsubsection{Availability of Source Code (\textbf{M$_4$})}
Our survey reveals that only 3 out of 16 publications provide access to their source code.
Most algorithms in literature study are intrinsically complex.
For example, among the algorithms \cite{yuan2023privgraph,ju2019generating,chen2019publishing,liu2020local} rely on \textit{community detection} \cite{nguyen2016detecting,mohamed2022differentially,fu2024community}, we find that minor differences in the implementation or parameters (e.g., allocating the privacy budget in each iteration) can have a significant impact on the overall utility.
Additionally, some open-sourced algorithms are implemented using different programming languages, such as \textit{Java} \cite{ye2020lf}, \textit{Python} \cite{yuan2023privgraph}, or \textit{C++} \cite{xiao2014differentially}, which makes it challenging to compare them fairly (i.e., efficiency issue).
Therefore, we encourage the public availability of implementations to provide additional insights and facilitate comparisons.

\begin{remark}
Most publications do not provide open-source codes, posing a significant challenge for the research community. 
Although a few algorithms \cite{xiao2014differentially,yuan2023privgraph} have available source codes, the lack of accessible codes for their competitors complicates the replication of experiments. 
\end{remark}

\begin{table*}[t]
\small
	\caption{Details of Graph Types in Different Algorithms.}
	\centering
	\label{tab:graph type}
		\begin{tabular}{|l|ccccccc|}
			\hline
	\diagbox[width=8.2 em]{Alg.}{Type} & Social (${\text{T}_1}$) & Web (${\text{T}_2}$) & Academic (${\text{T}_3}$) & Traffic (${\text{T}_4}$) & Financial (${\text{T}_5}$) & Technology (${\text{T}_6}$) & Synthetic (${\text{T}_7}$)  \\\hline

    DP-dK \cite{wang2013preserving} & \ding{51} & & \ding{51} & & & \ding{51} &  \\ 
    TmF \cite{nguyen2015differentially} & \ding{51} & \ding{51} & \ding{51} & & \ding{51} & &  \\ 
    DER \cite{chen2014correlated} & \ding{51} & & \ding{51} & \ding{51} & & & \\ 
    PrivSKG \cite{mir2012differentially} & & \ding{51} & \ding{51} & & & & \ding{51} \\ 
    PrivHRG \cite{xiao2014differentially} & \ding{51} & \ding{51} & \ding{51} & & & &  \\ 
    PrivGraph \cite{yuan2023privgraph} & \ding{51} & \ding{51} & \ding{51} & & & &  \\ 
    C-AGM \cite{chen2019publishing} & \ding{51} & \ding{51} & \ding{51} & & & &  \\ 
    TriCycLe \cite{jorgensen2016publishing} &  \ding{51} & \ding{51} & \ding{51} & & & &  \\ 
    PrivCom \cite{zhang2020community} &  \ding{51} & \ding{51} & \ding{51} & & & &  \\ 
    $\pi_v,\pi_e$ \cite{jian2021publishing} & \ding{51} &  & \ding{51} & & & & \ding{51} \\ 
    LDPGen \cite{qin2017generating} & \ding{51} &  &  & & & &  \\ 
    CGGen \cite{ju2019generating} & \ding{51} &  &  & & & &  \\ 
    LF-GDPR \cite{ye2020lf} &  \ding{51} &  &  \ding{51} & & & &  \\ 
    AsgLDP \cite{wei2020asgldp} & \ding{51} &  &  & & & &  \\ 
    Block-HRG \cite{hou2023block} & \ding{51} &  & \ding{51} & & & \ding{51} &  \\ 
    DP-LUSN \cite{liu2020local} &  & \ding{51} & \ding{51} & & &  &  \\
 \hline
	\end{tabular}
\begin{tablenotes}
     \item[1] Social (V: people, E: relationships) \quad Web (V: webpages, E: hyperlinks) \quad Academic (V: researchers, E: collaborations) \\
     \item[1]  Traffic (V: intersections, E: roads) \quad Financial (V: products, E: links) \quad Technology (V: apps, E: relationships)
\end{tablenotes}
\end{table*}

\subsection{Graph Datasets \textbf{G}}
Ideally, graph datasets used in the empirical analysis should consider the following key attributes (\textbf{G$_1$}$\sim$\textbf{G$_4$}): graph size (i.e., number of nodes or edges), average clustering coefficient (ACC), and graph types.

\subsubsection{Graph Size (\textbf{G$_1$}-\textbf{G$_2$})}
Our literature survey reveals that graph datasets used in different algorithms vary significantly in size, such as the number of nodes ($|\text{V}|$) and the number of edges ($|\text{E}|$). 
The size of graphs plays a crucial role in their utility and efficiency.
On the one hand, graph size determines the \textit{density}, which is an important metric for measuring the sparsity of graphs, represented as $\frac{2|\text{E}|}{{|\text{V}|}^2}$. 
Real-world graphs are usually sparse (low density), meaning that $|\text{E}|$ is much smaller than the maximum possible number of edges, i.e., $|\text{E}| \ll \frac{|\text{V}|(|\text{V}|-1)}{2}$. 
However, some perturbation mechanisms, such as randomized response, add significant noise to a graph, resulting in a much denser synthetic graph and undermining the utility \cite{ye2020lf, qin2017generating}. 
Theoretically, the sparser the graph, the more significant the density problem becomes.
On the other hand, processing time also increases with graph size. 
Therefore, to ensure the comprehensiveness of a benchmark, various graph datasets with different sizes should be evaluated.

\subsubsection{Average Clustering Coefficient (\textbf{G$_3$})}
The clustering coefficient \cite{holland1971transitivity} is a fundamental metric in graph theory, quantifying the extent to which nodes in a graph cluster together. 
This metric offers valuable insights into the local connectivity of the graph by indicating the probability that two neighbors of a given node are also neighbors of each other.
The clustering coefficient can be calculated: $C_i=e_i/\binom{{d_i}}{2}$, where $e_i$ is the number of edges in the subgraph of $G$ induced by a node $v_i$'s neighbors, and $d_i$ is node degree of $v_i$.

The average clustering coefficient (ACC) \cite{watts1998collective} measures the overall clustering within a network by averaging the clustering coefficients of all nodes. 
A network with a high ACC and a small average path length is often referred to as a "small-world" network.
The formal definition can be represented by:
    \begin{equation}
         \overline{C}=\frac{1}{n}\sum_{i=1}^{n}C_i=\frac{2}{n}\sum_{i=1}^{n}\frac{e_i}{d_i(d_i-1)},
    \end{equation}
where $n$ is the number of nodes in a graph.

Our literature survey indicates that graph datasets used in algorithms exhibit significant variation in ACC. 
Intuitively, some synthetic graph algorithms \cite{yuan2023privgraph,ju2019generating,chen2019publishing,liu2020local,qin2017generating} that leverage community or clustering information perform exceptionally well on graphs with high ACC. 
Therefore, a fair and comparable benchmark should include graphs with a range of ACCs.

\begin{table*}[t]
\small
	\caption{Graph Queries.}
	\centering
	\label{tab:graph query}
		\begin{tabular}{|l|ccc|ccc|ccc|ccccc|c|}
			\hline
	\multirow{3}{*}{\diagbox{Alg.}{Query}} & \multicolumn{3}{c|}{Counting} & \multicolumn{3}{c|}{Degree} & \multicolumn{3}{c|}{Path} & \multicolumn{5}{c|}{Topology} & Centrality\\
   \cline{2-16}
    & ${\text{Q}_1}$ & ${\text{Q}_2}$ & ${\text{Q}_3}$ & ${\text{Q}_4}$ & ${\text{Q}_5}$ & ${\text{Q}_6}$ & ${\text{Q}_7}$ & ${\text{Q}_8}$ & ${\text{Q}_9}$ & ${\text{Q}_{10}}$ & ${\text{Q}_{11}}$ & ${\text{Q}_{12}}$ & ${\text{Q}_{13}}$ & ${\text{Q}_{14}}$ & ${\text{Q}_{15}}$\\ \cline{2-16}
     & $|\text{V}|$  & $|\text{E}|$ & $\triangle$  & $\overline{d}$ & 
$d_{\sigma}$ & $\bm{d}$ & $l_{max}$ & $\overline{l}$ & $\bm{l}$ & GCC  & ACC & CD & Mod & Ass & EVC \\ 
\hline

    DP-dK \cite{wang2013preserving} & \ding{51} & \ding{51} & \ding{51}  & \ding{51} & & &\ding{51} & \ding{51} &  & & \ding{51} & &\ding{51} & &\ding{51} \\ 
    TmF \cite{nguyen2015differentially} & && & \ding{51} & \ding{51} & \ding{51} & \ding{51} & \ding{51} &  \ding{51} &  \ding{51} & & & &&\\ 
    DER \cite{chen2014correlated} &\ding{51} &&& & & \ding{51} & & \ding{51} & &&&&&& \\ 
    PrivSKG \cite{mir2012differentially} & &&& & & \ding{51} & &&&& \ding{51} && && \ding{51}\\ 
    PrivHRG \cite{xiao2014differentially} &  &&& & & \ding{51} & & & \ding{51} & &&&&& \ding{51} \\ 
    PrivGraph \cite{yuan2023privgraph} & &&& & & \ding{51} & \ding{51}& &  & \ding{51}& & \ding{51} & && \ding{51}\\ 
    C-AGM \cite{chen2019publishing} & & \ding{51} & \ding{51} & && \ding{51} & &&& \ding{51} & &&&&\\ 
    TriCycLe \cite{jorgensen2016publishing} & & \ding{51} & \ding{51} & && \ding{51} & &&& \ding{51} & \ding{51} & &&& \\ 
    PrivCom \cite{zhang2020community} & &  & & && & &&& & & \ding{51} &&&\\ 
    $\pi_v,\pi_e$ \cite{jian2021publishing} &  \ding{51} & \ding{51} & \ding{51} & && \ding{51} & &&& \ding{51} & \ding{51} &&&&\\ 
    LDPGen \cite{qin2017generating} &  &  & & && & &&& \ding{51} & & \ding{51} & \ding{51} &\ding{51}& \\ 
    CGGen \cite{ju2019generating} & &  & & && & &&& \ding{51} & &  & \ding{51} &\ding{51}& \\ 
    LF-GDPR \cite{ye2020lf} &  &  & & && & &&& \ding{51} & & \ding{51} & \ding{51} && \\ 
    AsgLDP \cite{wei2020asgldp} & &&&&& \ding{51} & &&& \ding{51} & & & \ding{51} &&\\ 
    Block-HRG \cite{hou2023block} &  &  & &\ding{51} && \ding{51}& &&& \ding{51}&\ding{51} & &\ding{51} &&\\ 
    DP-LUSN \cite{liu2020local} &  & \ding{51} & & && & &&& \ding{51} & & &&&\\ \hline
\end{tabular}
\begin{tablenotes}
        \item[1] \quad Table \ref{tab:symnbols of graph queries} provides details for each query symbol.
\end{tablenotes}
    \vspace{-10pt}
\end{table*}

\subsubsection{Graph Type (\textbf{G$_4$})}
As presented in Table \ref{tab:graph type}, multiple graphs from various domains are used to verify the performance of algorithms. 
Our literature review reveals that three graph types (social, web, and academic) are commonly used in most algorithms, while another three types (traffic, financial, and technology) are evaluated less frequently.
Graphs of different types possess distinct characteristics (e.g., node size, edge size, graph density, average clustering coefficient, number of triangles, etc.) that can influence the performance of proposed synthetic methods. 
For example, the social graphs often exhibit strong community structures, which is suitable for some community-based graph synthetic algorithms \cite{yuan2023privgraph,ju2019generating,chen2019publishing,liu2020local}. Therefore, it's important to consider a variety of graphs in experimental evaluations to have a fair assessment to algorithms' performance.

Additionally, the synthetic graph (${\text{T}_7}$) can simulate special characteristics that real-world graphs may not possess, such as binomial or uniform distributions. Although only two algorithms~\cite{mir2012differentially,jian2021publishing} evaluate synthetic graphs, as shown in Table \ref{tab:graph type}, we advocate for the inclusion of synthetic graphs in experiments to ensure the comprehensiveness of a benchmark.

\subsection{Privacy Requirements \textbf{P}}
In differentially private graph synthetic algorithms, data owners express their privacy requirements by controlling the privacy budget $\varepsilon$. 
In Table \ref{tab:comparisons of papers}, the range of privacy budgets in various publications differs significantly, ranging from 0.01 to 2000. 
In fact, using an excessively large $\varepsilon$ (e.g., 2000) could be meaningless for protecting information.
To facilitate the comparability of a benchmark, the privacy budget should be set reasonably and identically.
Additionally, some generation algorithms, such as DP-dK \cite{wang2013preserving}, PrivSKG \cite{mir2012differentially}, PrivCom \cite{zhang2020community}, provide ($\varepsilon,\delta$)-DP that is a relaxation of $\varepsilon$-DP.
It introduces an additional parameter $\delta$ to account for the allowable probability that the privacy guarantee may be violated.
In general, a randomized algorithm is considered safe when $\delta$ is preferably smaller than $1/n$ \cite{dwork2006our,abadi2016deep}, where n is the number of users.

\begin{table}[t]
\small
	\caption{Details of Graph Queries and Metrics.}
	\label{tab:symnbols of graph queries}
	\centering
		\begin{tabular}{|c|l|l|}
			\hline
               Query & \multicolumn{1}{c|}{Description} & \multicolumn{1}{c|}{Metrics} \\ \hline
                $|\text{V}|$ &  Number of nodes & RE, MRE \\ 
                $|\text{E}|$ & Number of edges &  RE, MRE \\
                $\triangle$  & Triangle counts & RE, MRE \\
                $\overline{d}$ & Average degree & RE \\
                $d_{\sigma}$ & Degree variance & RE \\ 
                $\bm{d}$ & Degree distribution & KL, HD, KS \\ 
                $l_{max}$ & Diameter & RE \\
                $\overline{l}$ & Average of all shortest paths & RE \\
                $\bm{l}$ & Distance distribution & RE \\ 
                GCC  & Global clustering coefficient & RE, MRE \\
                ACC & Average clustering coefficient & RE, MRE, MSE \\ 
                CD & Community detection & NMI, Avg-F$_1$, \\
                & & ARI, AMI\\
                Mod & Modularity & RE\ \\
                Ass & Assortativity coefficient & RE \\ 
                EVC & Eigenvector centrality & MAE \\ \hline
   \end{tabular}
    \begin{tablenotes}
     \item[1] RE (${\text{E}_1}$): relative error \quad MRE (${\text{E}_2}$): mean relative error; 
     \item[1] KL (${\text{E}_3}$): KL-divergence \quad HD (${\text{E}_4}$): Hellinger distance 
     \item[1] KS (${\text{E}_5}$): Kolmogorov-Smirnov statistic 
     \item[1] Avg-F$_1$ (${\text{E}_6}$): average F$_1$ score
     \item[1] MAE (${\text{E}_7}$): mean absolute error \quad MSE (${\text{E}_8}$): mean square error
     \item[1] ARI (${\text{E}_9}$): adjusted random index 
     \item[1] AMI (${\text{E}_{10}}$): adjusted mutual information
     \item[1] NMI (${\text{E}_{11}}$): normalized mutual information 
   \end{tablenotes}
\end{table}

\subsection{Utility \textbf{U}}
We consider two principles in the evaluation of generation algorithms: graph query (\textbf{U$_1$}) and error metric (\textbf{U$_2$}).

\subsubsection{Graph Query (\textbf{U$_1$})}
Multiple graph queries are employed to evaluate the performance of the proposed synthetic algorithms. 
As shown in Table \ref{tab:graph query}, we classify 15 graph queries into five categories: general counting, degree information, path condition, topology structure, and centrality. 
Table \ref{tab:symnbols of graph queries} provides the detailed content for each query.
Our literature survey indicates that all existing publications only cover a subset of these queries. 
In fact, some works evaluate only one of the five query types. For instance, LDPGen \cite{qin2017generating}, LF-GDPR \cite{ye2020lf}, and CGGen \cite{ju2019generating} focus solely on topology structure.
It is important to use a comprehensive set of graph queries to ensure a fair comparison of all algorithms.

\subsubsection{Error Metric (\textbf{U$_2$})}
For each graph query, researchers compare the error metric between the true and the noisy graph.
As illustrated in Table \ref{tab:symnbols of graph queries}, relative error (RE) is used to evaluate 12 out of 15 graph queries.
Given a query result of the true graph $Q(G)$ and a query result of the noisy graph $Q(G^\prime)$, RE can be computed as $\frac{|Q(G)-Q(G^\prime)|}{Q(G)}$.
Five graph queries (i.e., $|V|$, $|E|$, $\triangle$, GCC, and ACC) use the mean relative error (MRE) to calculate the utility loss, which can be represented as $\frac{1}{n}\sum_{i=1}^n |Q(G_i)-Q(G_i^\prime)|$, where $Q(G_i)$ (or $Q(G_i^\prime)$) is the result on the node $v_i$.
Additionally, some queries use special metrics to measure output results.
For instance, degree distribution is evaluated with Kullback-Leibler divergence (KL)~\cite{kullback1997information}, Hellinger distance (HD) \cite{nikulin2001hellinger}, or Kolmogorov-Smirnov statistic (KS) \cite{daniel1990kolmogorov}.
In community detection, the similarity of communities between the true and the synthetic graph can measured by normalized mutual information (NMI) \cite{kvalseth1987entropy}, average F$_1$ score \cite{zhang2020community,rossetti2017tiles}, adjusted random index (ARI) \cite{rand1971objective}, and adjusted mutual information (AMI) \cite{vinh2009information}.
Consistent use of error metrics in the benchmark is crucial for ensuring a fair comparison of all algorithms.

\section{Benchmark Instantiation}
\label{sec:benchmark instantiation}

In this section, we describe $\mathsf{PGB}$, a benchmark designed to evaluate the utility of differentially private synthetic graph algorithms. 
The goal of $\mathsf{PGB}$ is to establish a set of elements for empirical evaluation that satisfies the design principles outlined in Section \ref{sec:benchmark principle}. 
Table \ref{tab:benchmark instantiation} provides an overview of the $\mathsf{PGB}$ benchmark. 
Next, we discuss how each element meets the required criteria and how to maintain validity and comprehensiveness.

\subsection{Mechanisms \textbf{M}}
In this subsection, we discuss how to select algorithms in our benchmark to satisfy all design principles mentioned in Section \ref{sec:benchmark principle}.

\subsubsection{Mechanisms (\textbf{M$_1$}, \textbf{M$_2$}, and \textbf{M$_3$})}
As we discussed in Section\ref{sec:benchmark principle}, algorithms with different elements (i.e., privacy definition, sensitivity, (un)attributed) cannot be compared in a benchmark.
Instead, the graph synthesis algorithms included in the benchmark must adhere to the same privacy definition. 
Consistency in whether attributed information is protected should also be maintained. Besides, according to Table \ref{tab:comparisons of papers}, the edge CDP definition is employed in 8 out of 16 publications. 
Among them, 75\% of the algorithms target unattributed graph synthesis. 
Therefore, following the majority of publications, we evaluate unattributed graph
generation algorithms under edge CDP in $\mathsf{PGB}$. 
This can apply to  DP-dK\cite{wang2013preserving},  TmF \cite{nguyen2015differentially}, PrivSKG \cite{mir2012differentially}, PrivHRG\cite{xiao2014differentially}, PrivGraph\cite{yuan2023privgraph}, and DGG\cite{qin2017generating}.

\begin{remark}
  Our benchmark is not limited to edge CDP and unattributed graphs. 
  When the criteria are unified, any graph synthesis algorithms, such as those using edge LDP and attributed graphs, can be compared using this benchmark.
\end{remark}

\begin{table}[t]
\small
   \caption{PGB benchmark with 4-tuple (\textbf{M}, \textbf{G}, \textbf{P}, \textbf{U})}
	\label{tab:benchmark instantiation}
	\centering
		\begin{tabular}{|c|l|}
			\hline
               Element &  \multicolumn{1}{c|}{Instantiation} \\ \hline
        \textbf{M}  & (1) Model: Edge CDP \\ 
        &  (2) Unattributed graph \\ 
        &  (3) Algorithms:  DP-dK\cite{wang2013preserving},  TmF \cite{nguyen2015differentially}, PrivSKG \cite{mir2012differentially}, \\ 
        &  PrivHRG\cite{xiao2014differentially}, PrivGraph\cite{yuan2023privgraph}, DGG\cite{qin2017generating}\\
        \textbf{G} & 6 real-world graphs and 2 synthetic graphs (Table \ref{tab:graph datasets})\\
        \textbf{P} & $\varepsilon\in$ [0.1,10] \\
        \textbf{U} & 15 graph queries listed in Table \ref{tab:symnbols of graph queries}\\        
               \hline
   \end{tabular}
\end{table}

\subsubsection{Algorithm Implementation (\textbf{M$_4$})}
The correct implementation of algorithms is crucial to ensure the fairness and validity of empirical analysis. 
We implement algorithms in the benchmark based on the following principles:
(a) \textit{Original source code}. 
Unfortunately, this only holds for PrivHRG~\cite{xiao2014differentially} and PrivGraph \cite{yuan2023privgraph} (cf. Table \ref{tab:comparisons of papers}).
What's more, these two algorithms are implemented in different programming languages, namely, PrivHRG\footnote{\url{https://github.com/kaseyxiao/privHRG}} in C++ and PrivGraph\footnote{\url{https://github.com/Privacy-Graph/PrivGraph}} in Python.
(b) \textit{Reuse of components in SOTA algorithm}.
Multiple algorithms utilize the same components, such as graph queries, which can be applied across different algorithms. 
For instance, PrivGraph evaluates its performance using various graph queries (e.g., community detection, degree distribution, path condition), which are available as open-source tools. 
In such cases, we consistently apply these components across all algorithms.
(c) \textit{Correctness guarantee}.
We check the results of re-implemented algorithms to ensure that they align with the results reported in publications.
(d) \textit{Same programming language and running environment}.
To guarantee the fairness and validity of comparisons, we re-implement algorithms in Python and evaluate them in the same running environment.

As a result, we select six algorithms in our benchmark: DP-dK \cite{wang2013preserving}, TmF \cite{nguyen2015differentially}, PrivSKG \cite{mir2012differentially}, PrivHRG \cite{xiao2014differentially}, PrivGraph \cite{yuan2023privgraph}, and DGG \cite{qin2017generating}.
Among them, we use implementations from the authors for PrivGraph and PrivHRG, and re-implement other algorithms in Python.
It should be noted that we include one naive baseline DGG \cite{qin2017generating} mainly because it generates graphs based on node degrees, which are fundamental but significant pieces of features in differentially private graph algorithms \cite{ye2020lf, qin2017generating}. 
Since DGG is developed with LDP, we re-implement DGG with the central setting as our benchmark baselines.
All experiments are conducted on Linux machines running
Ubuntu 20.04.5 LTS with 16 AMD EPYC 7313P@3.7Ghz with 512GB of RAM.

\subsection{Graph Datasets \textbf{G}}
To meet the design principles outlined in Section \ref{sec:benchmark principle}, we conducted a series of benchmark experiments on a comprehensive set of graphs. 
Table \ref{tab:graph datasets} provides an overview of the graph datasets, summarizing four key properties: the number of nodes ($|\text{V}|$), the number of edges ($|\text{E}|$), the average clustering coefficient (ACC), and the graph types. 
The node sizes range from 2,600 to 22,687, and the edge sizes range from 3,300 to 250,278.
These graphs are sourced from seven different domains, with each graph type utilized at least once to evaluate a generation algorithm (cf. Table \ref{tab:graph type}). 
Among them, 6 out of 8 graphs are derived from public datasets (i.e., SNAP~\cite{snapnets}, NR\cite{nr}), while two are synthesized using generative models, specifically the Erdos-Renyi (ER) model~\cite{erd6s1960evolution} and the Barabasi-Albert (BA) model \cite{barabasi1999emergence}.
Node degrees in ER graphs follow a binomial distribution \cite{newman2001random}, whereas node degrees in BA graphs follow a power-law distribution \cite{bollobas2001degree}.
In our experiments, both the ER and BA graphs were generated with $|\text{V}|$ = 10,000.

\begin{table}[t]
\small
\caption{Details of Graph Datasets.}
\label{tab:graph datasets}
\centering
\begin{threeparttable}
    \begin{tabular}{|lrrrl|}
        \hline
       \multicolumn{1}{|c|}{Graph}  & \multicolumn{1}{c|}{$|\text{V}|$}    & \multicolumn{1}{c|}{$|\text{E}|$} & \multicolumn{1}{c|}{ACC}  & \multicolumn{1}{c|}{Type}  \\ \hline
        Minnesota\tablefootnote{\url{https://networkrepository.com/road-minnesota.php}} & 2,600 & 3,300 & 0.0160 & Traffic \\
        Facebook\tablefootnote{\url{http://snap.stanford.edu/data/ego-Facebook.html}} & 4,039 & 88,234 & 0.6055 & Social  \\ 
        Wiki-Vote\tablefootnote{\url{http://snap.stanford.edu/data/wiki-Vote.html}} & 7,115 & 103,689 & 0.1409 & Web  \\ 
        ca-HepPh\tablefootnote{\url{http://snap.stanford.edu/data/ca-HepPh.html}} & 12,008 & 118,521 & 0.6115 & Academic \\
        poli-large\tablefootnote{\url{https://networkrepository.com/econ-poli-large.php}} & 15,600 & 17,500 & 0.3967 & Financial \\
        Gnutella\tablefootnote{\url{http://snap.stanford.edu/data/p2p-Gnutella25.html}} & 22,687 & 54,705 & 0.0053 & Technology \\
        ER graph & 10,000 & 250,278 & 0.0050 & Synthetic \\
        BA graph & 10,000 & 49,975 & 0.0074 & Synthetic\\
        \hline
    \end{tabular}
\end{threeparttable}
\end{table}

\subsection{Privacy Requirements \textbf{P}}
Following the example of most experimental analyses in publications, we also conduct experiments with varying $\varepsilon$ values.
Determining an appropriate $\varepsilon$ is an ongoing area of research \cite{dwork2019differential,kasiviswanathan2014semantics,lee2011much,pankova2022interpreting}. 
In our experiments, we vary the privacy budget $\varepsilon$ from 0.1 to 10, similar to the ranges used in most studies listed in Table \ref{tab:comparisons of papers}.
For algorithms we implement in \textbf{M$_4$}, DP-dK \cite{wang2013preserving} and PrivSKG \cite{mir2012differentially} maintain ($\varepsilon,\delta$)-DP, while the others provide $\varepsilon$-DP.
To ensure a fair comparison, we set $\delta$ = 0.01 for DP-dK and PrivSKG, following the parameters used in this work\cite{wang2013preserving,mir2012differentially}.

\subsection{Utility \textbf{U}}
To ensure the comparability of our benchmark, we apply all the queries listed in Table \ref{tab:graph query} to evaluate the performance of the algorithms. 
These queries represent the union of those used in 16 different publications. 
Due to the inherent randomness of the algorithms, the utility can differ significantly under the same combination of privacy budget and graph dataset.
Similar to various studies in related work, we run each experiment 10 times and calculate the average of the utility metrics. 
We use different metrics for various graph queries. 
First, we use Relative Error (RE) for most queries, including $|\text{V}|$, $|\text{E}|$, $\triangle$, $\overline{d}$, $d_{\sigma}$, $l_{max}$, $\overline{l}$, GCC, ACC, Mod, and Ass. 
Second, we use Kullback-Leibler divergence (KL), Normalized Mutual Information (NMI), and Mean Absolute Error (MAE) to evaluate the utility error of $\bm{d}$, CD, and EVC, respectively. 
Third, we use KL for $\bm{l}$ instead of RE, as KL can better measure how one probability distribution differs from another compared to RE.
The details of graph queries and metrics are explained in Table~\ref{tab:symnbols of graph queries}.

\section{Experimental Results}
\label{sec:experiment}

We formulate the following research questions:
\begin{itemize}
  \item \textbf{Q1}: 
  How do algorithms compare in terms of the overall utility across various graphs and privacy budgets?
  \item \textbf{Q2}: 
   How do graph datasets, privacy budgets, and utility metrics affect the utility of different algorithms?
  \item \textbf{Q3}: What are the time and space costs of the algorithms?
\end{itemize}

\subsection{Overall Utility Analysis}

\begin{table*}[t]
\small
   \caption{Overall Results.}
	\label{tab:overall result}
	\centering
 	\setlength{\tabcolsep}{5mm}{
    \begin{threeparttable}
       \resizebox{2\columnwidth}{!}{
		\begin{tabular}{|c|c|c|c|c|c|c|c|c|c|}
			\hline
    $\varepsilon$ & Algorithms & \multicolumn{8}{c|}{Graph Datasets} \\ \cline{3-10}
    & & Minnesota & Facebook & Wiki &  HepPh & Poli & Gnutella & ER & BA \\ \hline
    0.1 & DP-dK & 5 & 4 & 3 &3 &4  & 2& 0 & 0 \\
    & TmF       & \cellcolor{gray!50}6 & 4 & 3 & 3 & \cellcolor{gray!50}5 & \cellcolor{gray!50}4& \cellcolor{gray!50}14&\cellcolor{gray!50} 6 \\
    & PrivSKG   & 1 & 1 & 3 & 2& 2 & 3& 2 &2 \\
    & PrivHRG   & 2 & 0 & 1 &0 & 2 & \cellcolor{gray!50}4& 2 & 3\\
    & PrivGraph & 1 & 1 & 1 &2 & 2 & 1& 1 &3 \\
    & DGG       & 2 & \cellcolor{gray!50}7 & \cellcolor{gray!50}6 & \cellcolor{gray!50}7& 2 &3 & 1 &3 \\  \hline
    0.5 & DP-dK & \cellcolor{gray!50}5 & 5 & 1 &4 & 2 &2 & 0 & 1\\
    & TmF       & \cellcolor{gray!50}5 & 4 & 3 &3 & \cellcolor{gray!50}4 & \cellcolor{gray!50}5& \cellcolor{gray!50}13 & 4\\
    & PrivSKG   & 2 & 0 & 3 &2 & 3 &4 & 2 &\cellcolor{gray!50}6 \\
    & PrivHRG   & 2 & 0 & 2 &0 & 3 &3 & 3 & 3\\
    & PrivGraph & 2 & 2 & 1 &1 & 1 & 1& 1 & 1\\
    & DGG       & 1 & \cellcolor{gray!50}7 &\cellcolor{gray!50} 7 &\cellcolor{gray!50}7 & \cellcolor{gray!50}4 &2 & 1 &2 \\  \hline
    1 & DP-dK  & \cellcolor{gray!50}5 &\cellcolor{gray!50} 5 & 2 & 3& 2 &2 & 0 & 1\\
    & TmF       & 4 & 4 & 3 &3 & \cellcolor{gray!50}4 &\cellcolor{gray!50}5 & \cellcolor{gray!50}12 & 4\\
    & PrivSKG   & 3 & 0 & \cellcolor{gray!50}4 &2 & 2 &5 & 2 &4 \\
    & PrivHRG   & 1 & 0 & 2 &0 & 3 &1 & 4 & 1\\
    & PrivGraph & 3 & 2 & 2 &4 & 2 &2 & 1 & \cellcolor{gray!50}5\\
    & DGG       & 1 & \cellcolor{gray!50}6 & \cellcolor{gray!50}4 &\cellcolor{gray!50} 5& \cellcolor{gray!50}4 &2 & 1 & 2\\  \hline
    2 & DP-dK    & \cellcolor{gray!50}4 &\cellcolor{gray!50} 6 & 2 & \cellcolor{gray!50}5& 2 & 3& 0 &1 \\
    & TmF       & 3 & 4 & 3 &3 & \cellcolor{gray!50}4 & \cellcolor{gray!50}4& \cellcolor{gray!50}13 & 4\\
    & PrivSKG   & \cellcolor{gray!50}4 & 0 & 2 & 2& 3 & \cellcolor{gray!50}4& 2 &\cellcolor{gray!50} 8\\
    & PrivHRG   & 0 & 0 & 2 & 0& 2 & 1& 3 &1 \\
    & PrivGraph & \cellcolor{gray!50}4 & 2 & \cellcolor{gray!50}4 &3 & 2 &2 & 1 & 1\\
    & DGG       & 2 & 5 & \cellcolor{gray!50}4 &4 & \cellcolor{gray!50}4 & 3& 1 & 2\\  \hline
    5 & DP-dK   & 4 & 5 & 2 &\cellcolor{gray!50}6 & 2 &3 & 1 & 1\\
    & TmF       & 4 & 5 & 4 & 4& \cellcolor{gray!50} 4& \cellcolor{gray!50}4& \cellcolor{gray!50}11 & 3\\
    & PrivSKG   & \cellcolor{gray!50}5 & 0 & 1 & 1& 3 & \cellcolor{gray!50}4& 2 & \cellcolor{gray!50}7\\
    & PrivHRG   & 0 & 0 & 1 &0 & 3 &1 & 4 &2 \\
    & PrivGraph & 2 & 2 & \cellcolor{gray!50}6 & 2& 2 &2 & 1 &2 \\
    & DGG       & 2 &\cellcolor{gray!50} 6 & 3 & 4& 3 & 3& 1 & 2\\  \hline
    10 & DP-dK  & 4 & \cellcolor{gray!50}5 & 1 & 4& 2 &3 & 1 & 1\\
    & TmF       & \cellcolor{gray!50}8 & \cellcolor{gray!50}5 & \cellcolor{gray!50}11 &\cellcolor{gray!50}9 & \cellcolor{gray!50}4 &\cellcolor{gray!50}8 & \cellcolor{gray!50}13 & \cellcolor{gray!50}8\\
    & PrivSKG   & 0 & 0 & 0 & 1& 3 & 3& 2 & 3\\
    & PrivHRG   & 0 & 0 & 2 &0 & 3 & 0& 2 & 2\\
    & PrivGraph & 2 & 3 & 1 & 1& 2 & 1& 1 & 1\\
    & DGG       & 3 & 4 & 2 & 2& 3 & 2& 1 & 2\\  \hline
   \end{tabular}
   }
    \begin{tablenotes}
    \item[1] Each number shows how often the algorithm performs best across 15 queries, given a privacy budget $\varepsilon$ and a graph dataset. 
    For example, the first number '5' means that DP-dK outperforms others in 5 queries (i.e., ${\text{Q}_5}$, ${\text{Q}_6}$, ${\text{Q}_9}$, ${\text{Q}_{12}}$, ${\text{Q}_{13}}$) for the Minnesota graph with $\varepsilon = 0.1$.
    \item[2] The highest frequency in each case is highlighted in gray.
   \end{tablenotes}
   \end{threeparttable}
   }
\end{table*}

We first present the comprehensive results of our benchmark study on differentially private graph generation algorithms. 
Table \ref{tab:overall result} summarizes the performance of six state-of-the-art algorithms across various graph datasets under different privacy budgets ($\varepsilon$). 
Each entry in the table indicates the number of times an algorithm achieved the best performance out of 15 queries for a given dataset and privacy budget (Definition~\ref{def:overall metric}). 
The highest frequency in each case is highlighted in gray.
We can conclude some key findings from the overall results.

\begin{definition}
\label{def:overall metric}
Let $\mathcal{A}$ be target algorithm.
Let $G$ and $\varepsilon$ be the graph dataset and privacy budget, respectively.
Let \(\mathcal{Q} = \{Q_1, Q_2, \ldots, Q_{p}\}\) be a set of $p$ queries. 
Let \( B_{i} \)  be the best performance indicator:
   \[
   B_{i} = 
   \begin{cases} 
   1 & \text{if } A \text{ performs best on } Q_i \text{ for } G \text{ and } \varepsilon \\
   0 & \text{otherwise}
   \end{cases}
   \]
Finally, we have:
   \[
   C_A(G,\varepsilon) = \sum_{i=1}^{p} B_{i},
   \]
where \( C_A(G,\varepsilon)  \) is the count of how often algorithm \( A \) performs best across the $p$ queries for \( G \) and \(\varepsilon\).
\end{definition}


\textit{Impact of Graph Dataset}:
We evaluate the performance of different algorithms on multiple graph datasets with various characteristics (e.g., sizes, ACC values, and types).
We have the following observations from Table \ref{tab:overall result}.
1) Graph Size. 
DGG performs well on the graph datasets with small size (i.e., $|\text{V}|<10^4$).
The reason is that DGG randomly generates intra-cluster edges according to the degree information, which is susceptible to graph size.
TmF behaves better than other methods when the graph size become larger (i.e., $|\text{V}| \geq 10^4$).
TmF perturbs the adjacency matrix directly, which preserves the structure information to some extent.
2) ACC. 
DGG performs better than other methods on graphs with high ACC values.
It is because DGG uses BTER algorithm to generate a synthetic graph and thus nodes with similar degrees are clustered together.
3) Graph Type. 
TmF performs well on multiple graph datasets from different domains, including real-world graphs and synthetic graphs.
It adds Laplace noise into each cell of the adjacency matrix, which is suitable for most of graph queries.

\textit{Impact of Privacy Budget}:
We compare different methods under a wide range of privacy budgets, with the following observations drawn from the results in Table \ref{tab:overall result}.
1) As the privacy budget $\varepsilon$ increases, TmF generally improves in performance. 
For example, TmF achieves top performance in 8 instances at $\varepsilon = 10$, the highest count in the entire table. 
TmF applies Laplace noise directly to each element of the adjacency matrix. 
As the privacy budget increases, less noise is added, allowing for better preservation of key information.
2) At lower privacy budgets (e.g., $\varepsilon = 0.1$), algorithm performance varies widely. 
No single method consistently dominates across all datasets, highlighting the complexity of achieving strong performance under strict privacy constraints. 
DP-dK and DGG outperform other methods when the privacy budget is small, as they generate synthetic graphs based on perturbed degree information, which is effective for most queries.
3) TmF achieves the most instances of top performance at both very low and very high privacy budgets on the Minnesota dataset, i.e., $\varepsilon=0.1$ and 10. 
However, it performs only moderately well at mid-range privacy budgets, i.e., $\varepsilon=1$, 2, and 5. 
This is because TmF outperforms other methods for queries ${\text{Q}_2}, {\text{Q}_7}$, and ${\text{Q}_8}$ when $\varepsilon=1$ and 10, but the result is reversed at $\varepsilon=0.5$, 1, 2, and 5. 
This indicates TmF’s performance variability in querying the number of edges, diameter, and shortest path.
4) PrivGraph achieves top performance in 4 and 6 instances on the Wiki dataset when $\varepsilon$=2 and 5, respectively. PrivGraph’s strength lies in accounting for different connection characteristics within and between communities, making it effective for querying distance distribution, global clustering coefficient, and modularity. However, it only achieves top performance 1 or 2 times at $\varepsilon$=0.1, 0.5, 1, and 10. At smaller privacy budgets (e.g., $\varepsilon=0.1$ or 0.5), PrivGraph introduces significant noise into community information, impacting accuracy. 
Conversely, at larger privacy budgets (e.g., $\varepsilon=10$), TmF outperforms PrivGraph by reducing the impact of noise on the adjacency matrix using a high-pass filter.

\textit{Overall Best Performers}:
According to results in Table \ref{tab:overall result}, we have the following observations.
1) TmF stands out as the most reliable and versatile algorithm across different privacy budgets and datasets.
The reason is that TmF leverages the high-pass filtering technique to avoid the whole matrix manipulation.
Nevertheless, when the privacy budget is small (i.e., $\varepsilon \leq 1$), other methods (i.e., DP-dK, PrivSKG, and DGG) perform better than TmF.
This is because TmF adds more noise into the elements of the matrix when $\varepsilon$ is small.
2) DGG emerges as a strong contender, particularly excelling in specific cases.
For instance, when $\varepsilon \leq 1$, DGG performs well on Facebook, Wiki-Vote, and ca-HepPh.
The reason is that DGG generates synthetic graphs based on degree information, which is vital for most graph queries.


\subsection{Utility in Specific Cases}

\begin{figure*}[h]
	\centering 
 	\subfigure{
		\begin{minipage}[t]{\linewidth}
			\centering
			\includegraphics[width=0.65\linewidth]{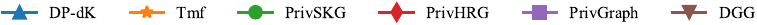}
               \vspace{-22pt}
		\end{minipage}
	}%
  \quad
	\subfigure{
		\begin{minipage}[t]{0.256\linewidth}
			\centering
			\includegraphics[width=\linewidth]{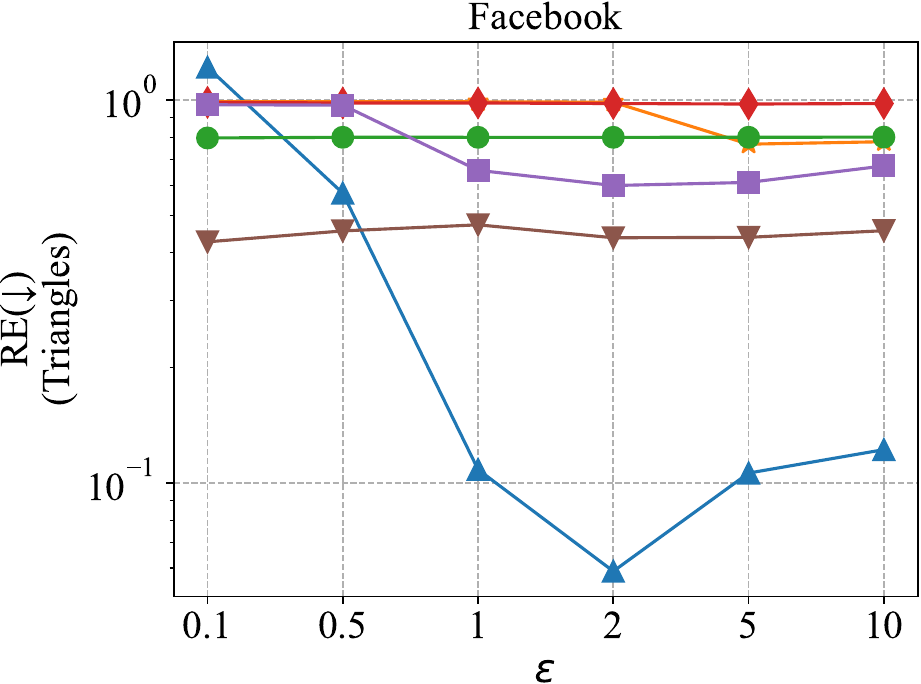}
			\label{fig:facebook triangles}
            \vspace{-15pt}
		\end{minipage}
	}%
	\subfigure{
		\begin{minipage}[t]{0.23\linewidth}
			\centering
			\includegraphics[width=\linewidth]{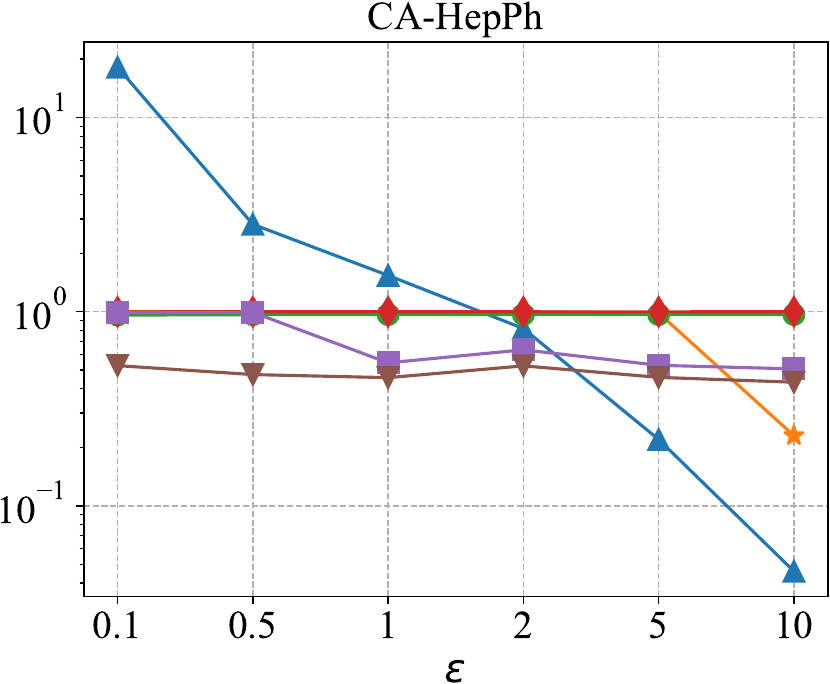}
			\label{fig:hepph triangles}
               \vspace{-15pt}
		\end{minipage}%
	}
	\subfigure{
		\begin{minipage}[t]{0.23\linewidth}
			\centering
			\includegraphics[width=\linewidth]{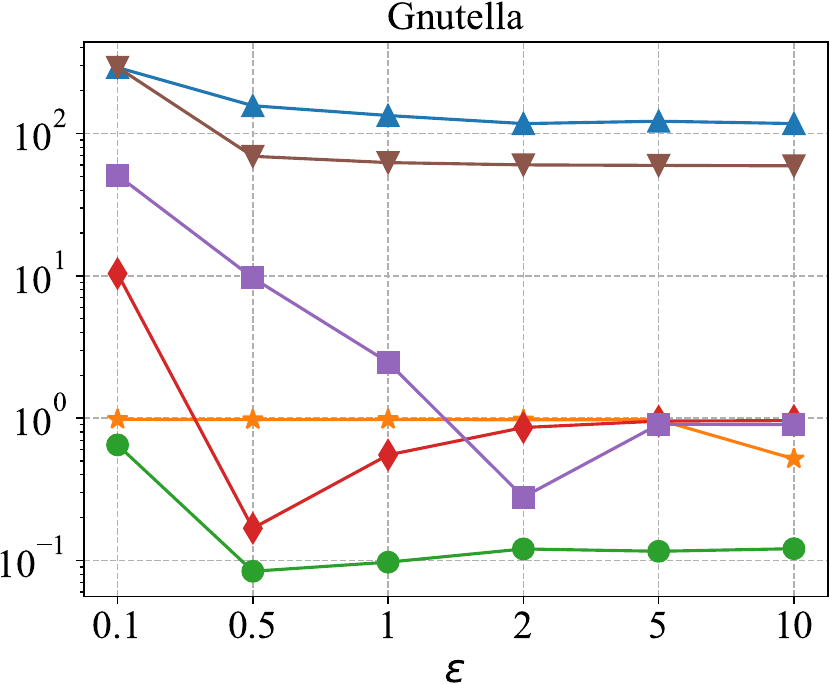}
			\label{fig:gnutella triangles}
               \vspace{-15pt}
		\end{minipage}%
	}
	\subfigure{
		\begin{minipage}[t]{0.23\linewidth}
			\centering
			\includegraphics[width=\linewidth]{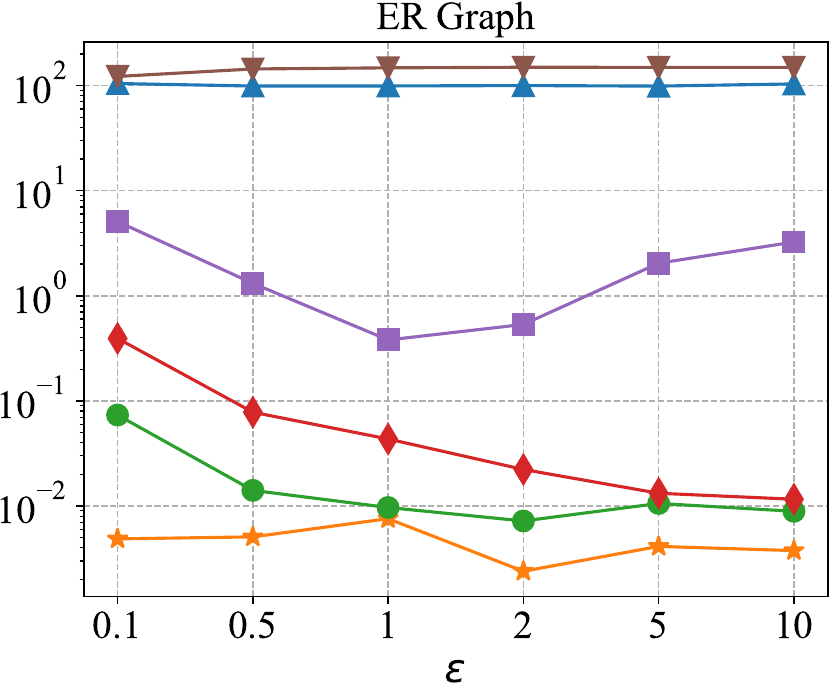}
			\label{fig:er triangles}
               \vspace{-15pt}
		\end{minipage}%
	}
	\qquad
	\subfigure{
		\begin{minipage}[t]{0.256\linewidth}
			\centering
			\includegraphics[width=\linewidth]{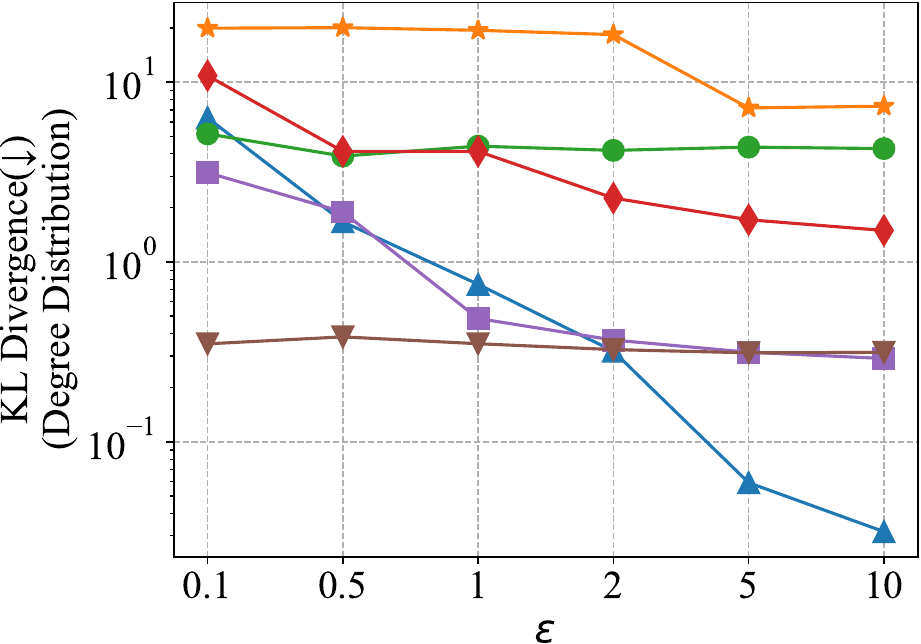}
			\label{fig:facebook degree}
               \vspace{-15pt}
		\end{minipage}
	}%
	\subfigure{
		\begin{minipage}[t]{0.23\linewidth}
			\centering
			\includegraphics[width=\linewidth]{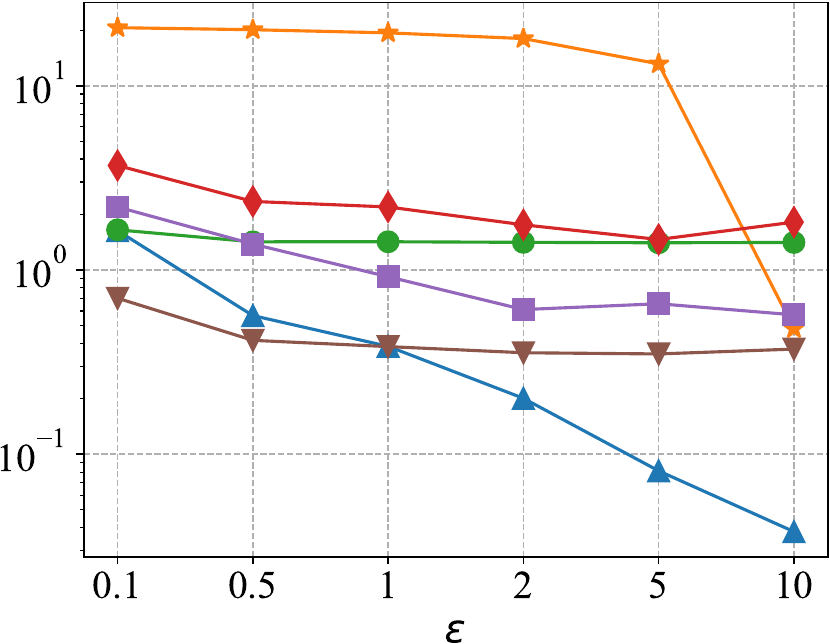}
			\label{fig:hepph degree}
               \vspace{-15pt}
		\end{minipage}%
	}
	\subfigure{
		\begin{minipage}[t]{0.23\linewidth}
			\centering
			\includegraphics[width=\linewidth]{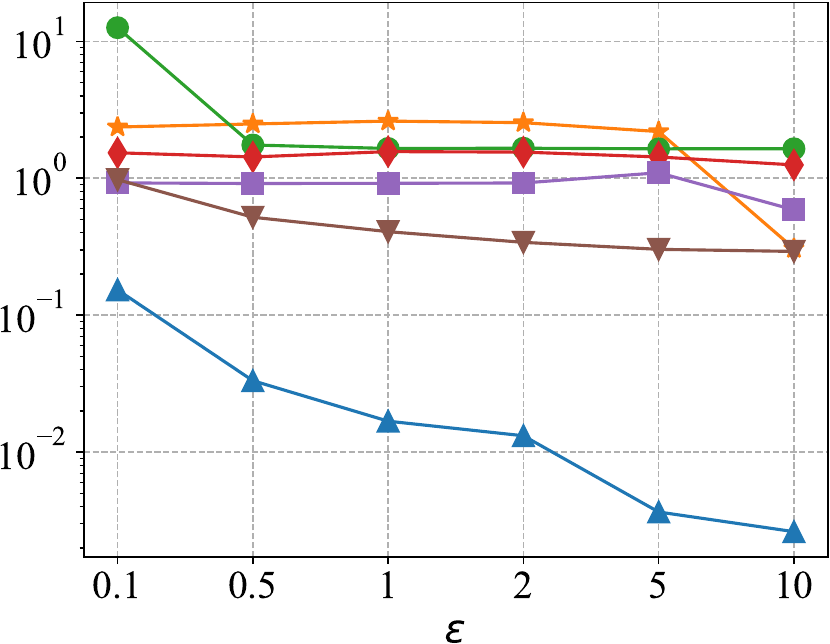}
			\label{fig:gnutellar degree}
               \vspace{-15pt}
		\end{minipage}%
	}
	\subfigure{
		\begin{minipage}[t]{0.23\linewidth}
			\centering
			\includegraphics[width=\linewidth]{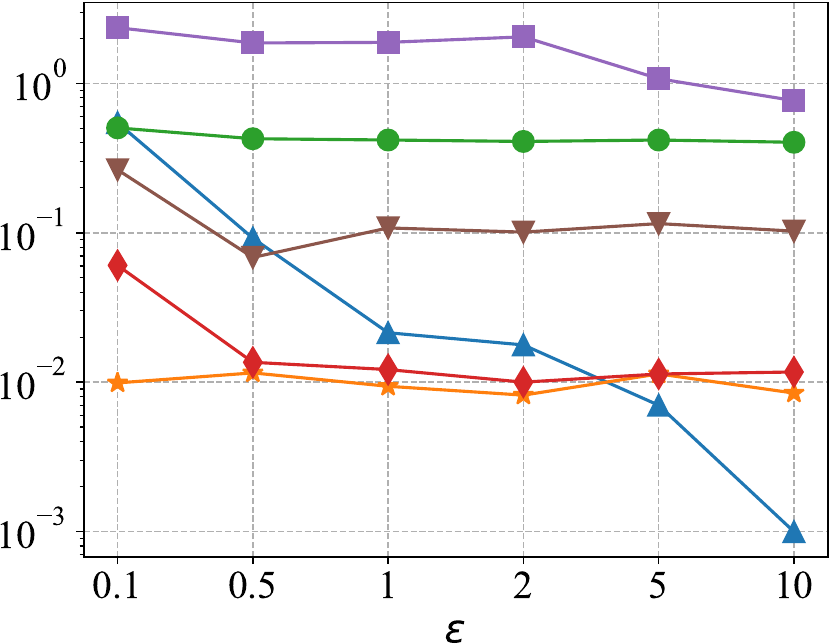}
			\label{fig:er degree}
               \vspace{-15pt}
		\end{minipage}%
	}
 	\qquad
	\subfigure{
		\begin{minipage}[t]{0.256\linewidth}
			\centering
			\includegraphics[width=\linewidth]{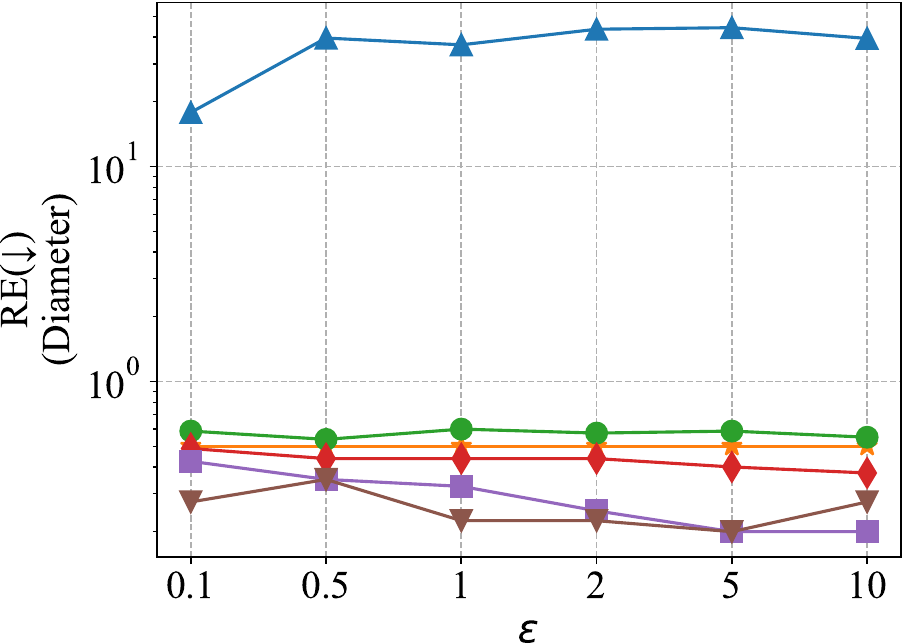}
			\label{fig:facebook diameter}
            \vspace{-15pt}
		\end{minipage}
	}%
	\subfigure{
		\begin{minipage}[t]{0.23\linewidth}
			\centering
			\includegraphics[width=\linewidth]{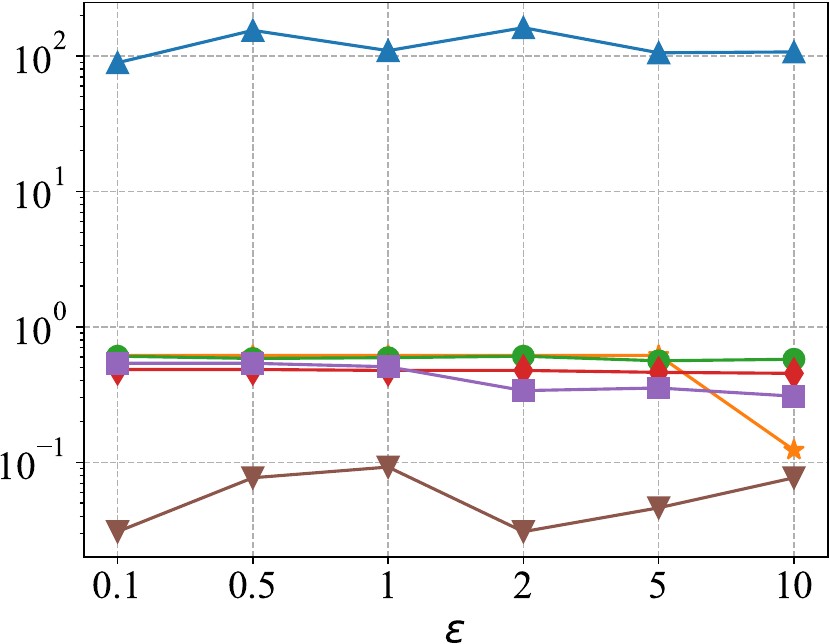}
			\label{fig:hepph diameter}
               \vspace{-15pt}
		\end{minipage}%
	}
	\subfigure{
		\begin{minipage}[t]{0.23\linewidth}
			\centering
			\includegraphics[width=\linewidth]{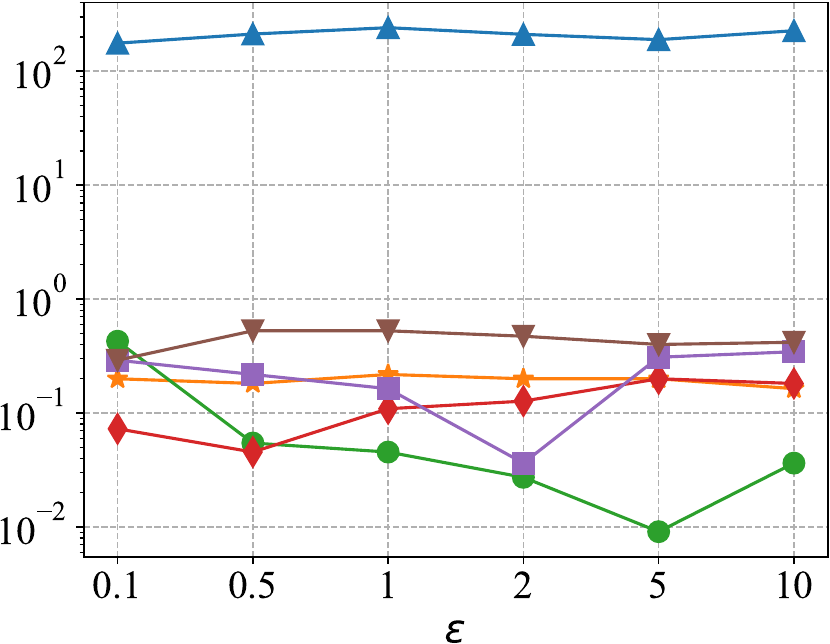}
			\label{fig:gnutellar diameter}
               \vspace{-15pt}
		\end{minipage}%
	}
	\subfigure{
		\begin{minipage}[t]{0.23\linewidth}
			\centering
			\includegraphics[width=\linewidth]{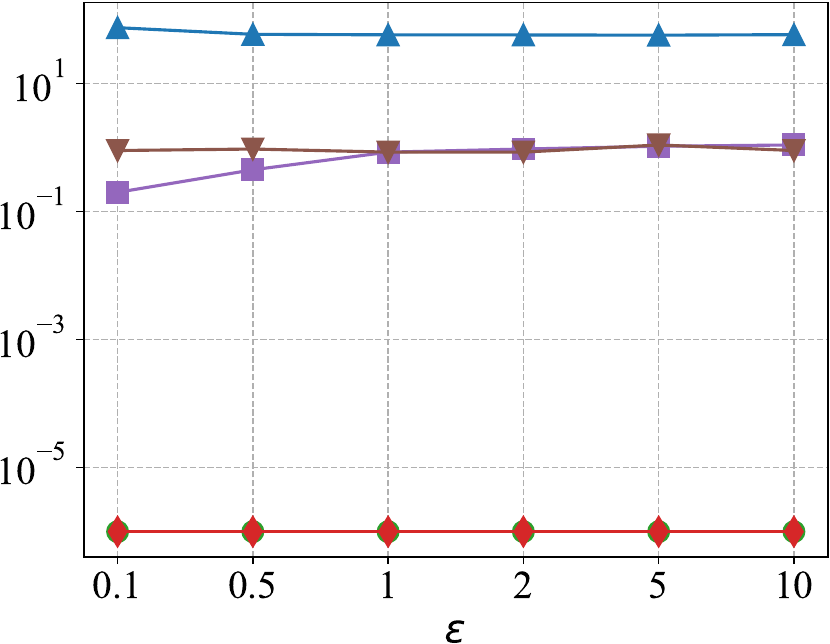}
			\label{fig:er diameter}
               \vspace{-15pt}
		\end{minipage}%
	}
   	\qquad
	\subfigure{
		\begin{minipage}[t]{0.256\linewidth}
			\centering
			\includegraphics[width=\linewidth]{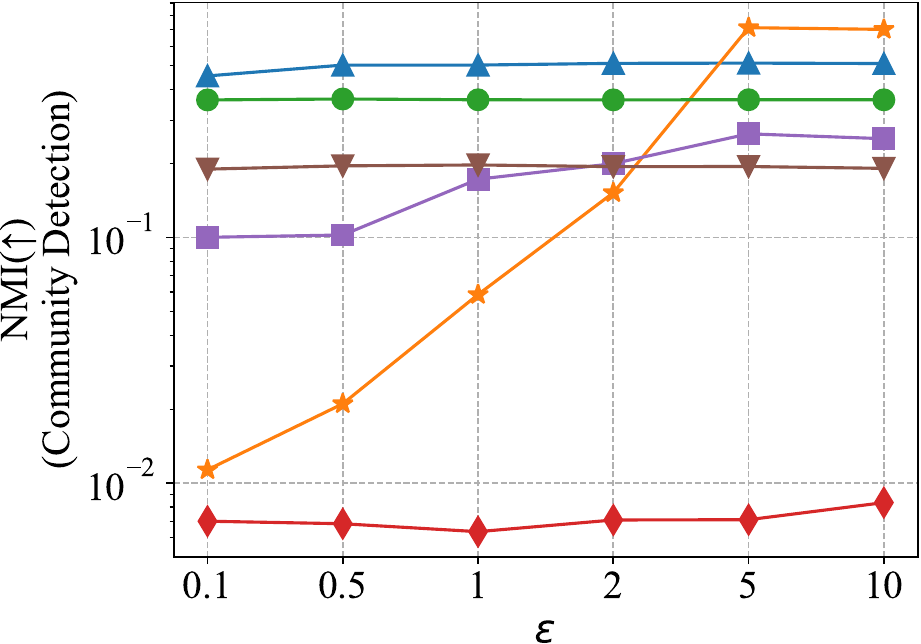}
			\label{fig:facebook cd}
               \vspace{-15pt}
		\end{minipage}
	}%
	\subfigure{
		\begin{minipage}[t]{0.23\linewidth}
			\centering
			\includegraphics[width=\linewidth]{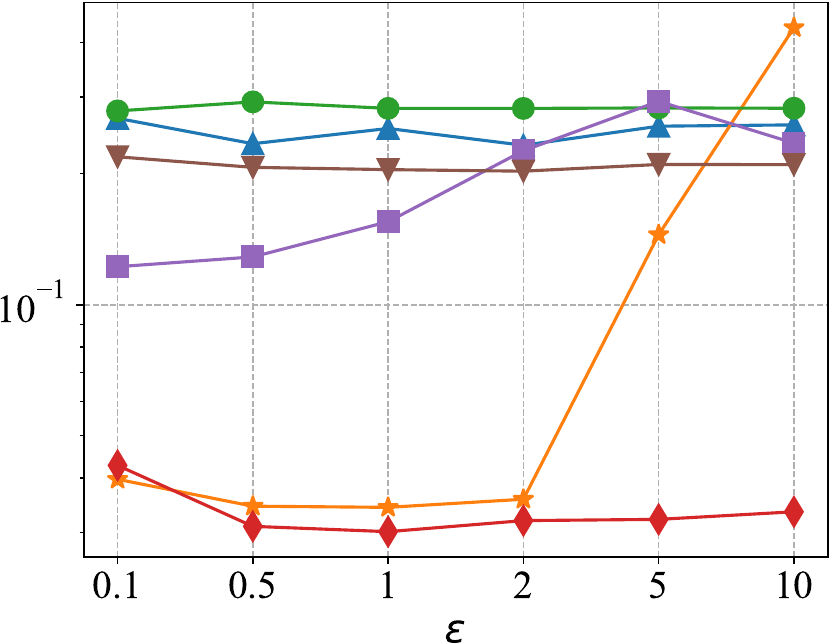}
			\label{fig:hepph cd}
               \vspace{-15pt}
		\end{minipage}%
	}
	\subfigure{
		\begin{minipage}[t]{0.23\linewidth}
			\centering
			\includegraphics[width=\linewidth]{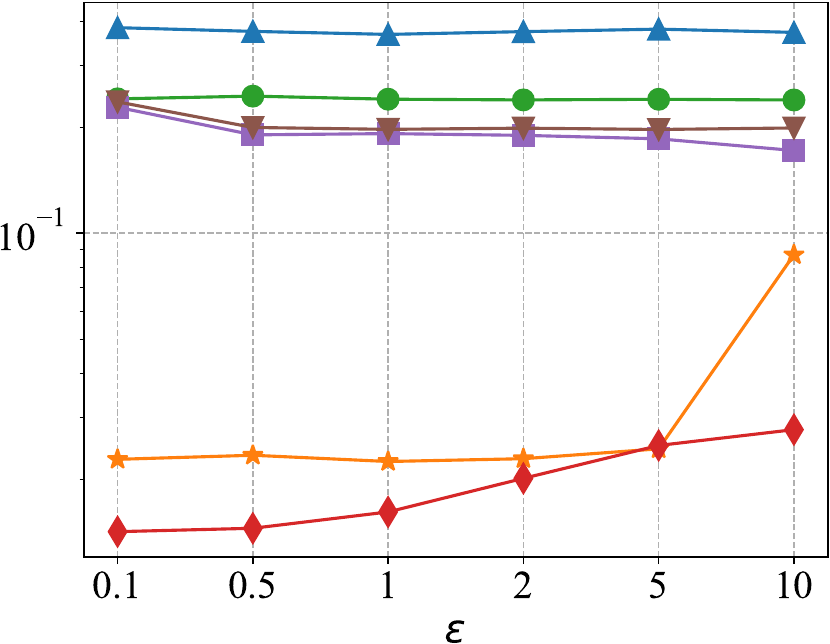}
			\label{fig:gnutellar cd}
               \vspace{-15pt}
		\end{minipage}%
	}
	\subfigure{
		\begin{minipage}[t]{0.23\linewidth}
			\centering
			\includegraphics[width=\linewidth]{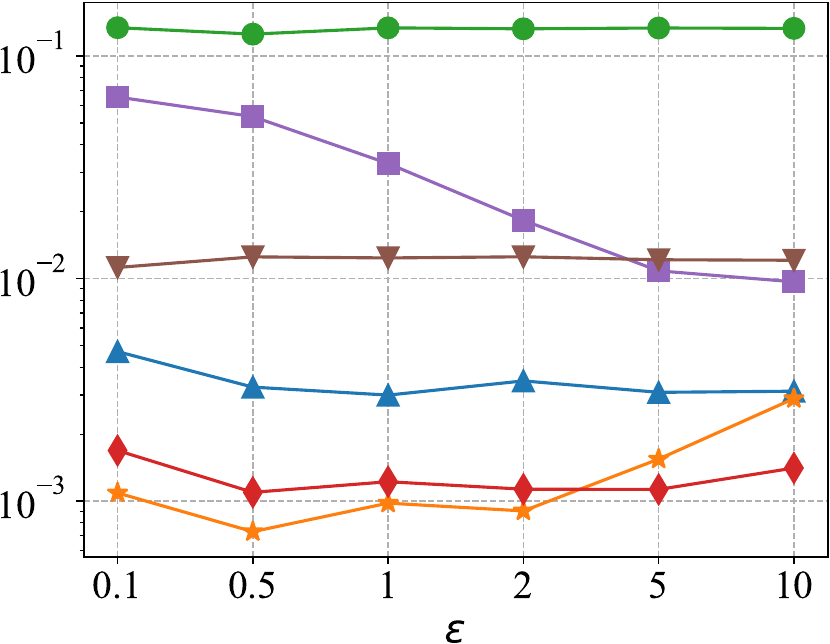}
			\label{fig:er cd}
               \vspace{-15pt}
		\end{minipage}%
	}
    \qquad
	\subfigure{
		\begin{minipage}[t]{0.256\linewidth}
			\centering
			\includegraphics[width=\linewidth]{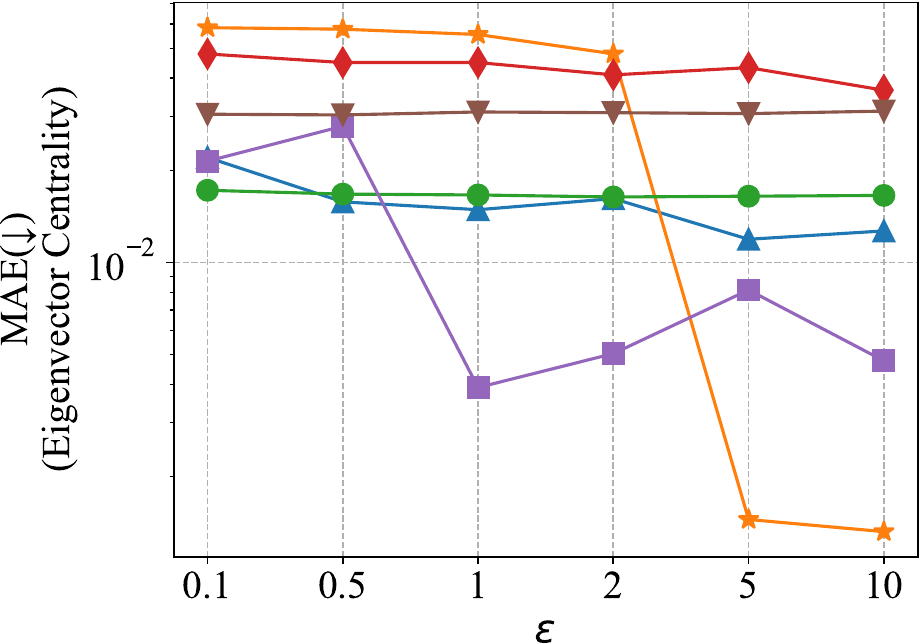}
			\label{fig:facebook evc}
               \vspace{-15pt}
		\end{minipage}
	}%
	\subfigure{
		\begin{minipage}[t]{0.23\linewidth}
			\centering
			\includegraphics[width=\linewidth]{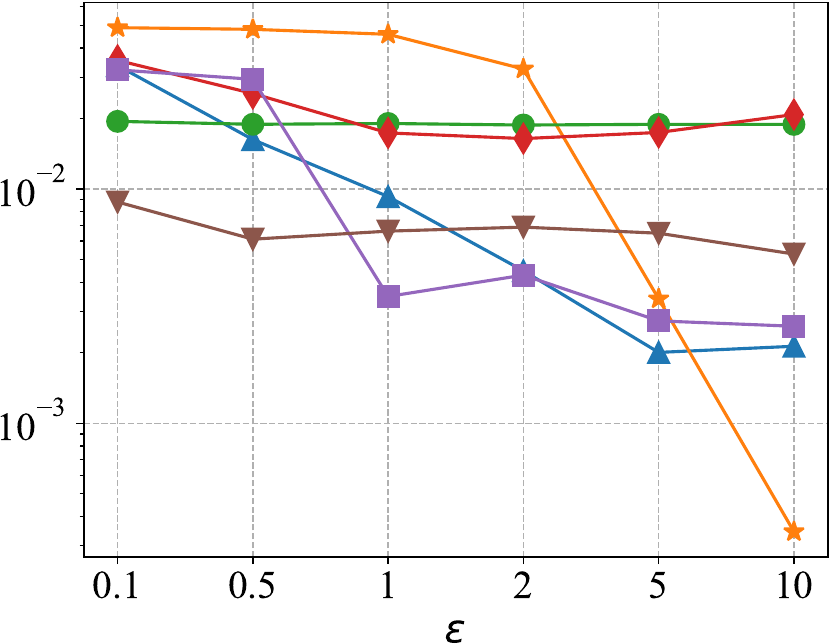}
			\label{fig:hepph evc}
               \vspace{-15pt}
		\end{minipage}%
	}
	\subfigure{
		\begin{minipage}[t]{0.23\linewidth}
			\centering
			\includegraphics[width=\linewidth]{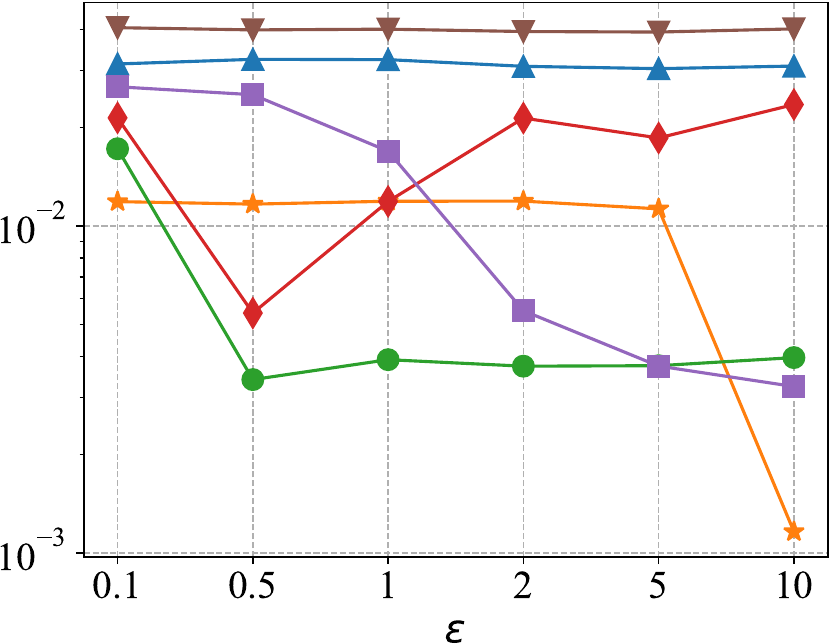}
			\label{fig:gnutellar evc}
               \vspace{-15pt}
		\end{minipage}%
	}
	\subfigure{
		\begin{minipage}[t]{0.23\linewidth}
			\centering
			\includegraphics[width=\linewidth]{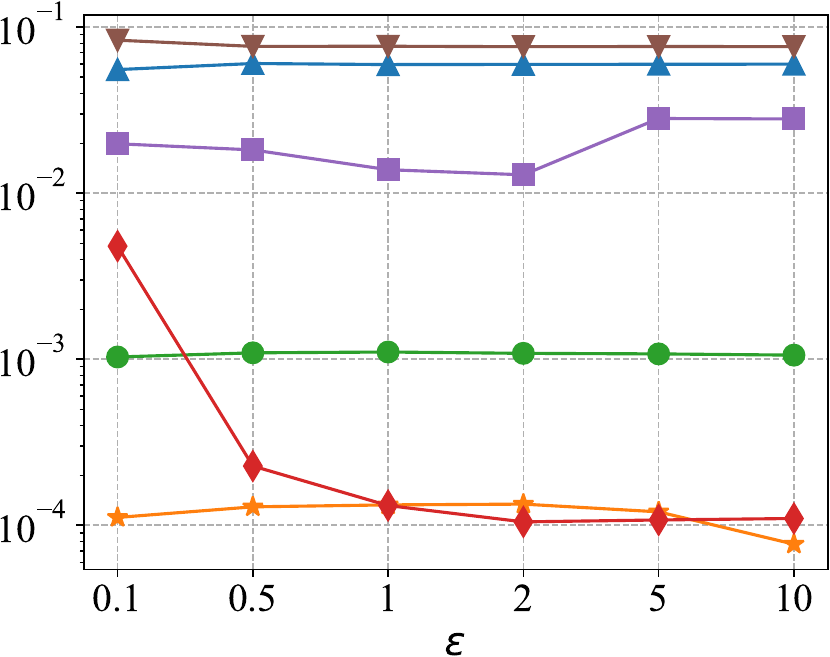}
			\label{fig:er evc}
               \vspace{-15pt}
		\end{minipage}%
	}
\centering
\vspace{-10pt}
\caption{End-to-end comparison of algorithms under different graph datasets, privacy budget $\varepsilon$ and qureies.
}
\label{fig:specific experimental results}
\vspace{-8pt}
\end{figure*}

To further illustrate the utility of algorithms, we examine specific cases from the benchmark results shown in Fig. \ref{fig:specific experimental results}.
Due to the limited space, we list the results of five queries on four graphs. 
The entire results of all cases can be accessed~\footnote{PGB:\url{https://github.com/dooohow/PGB}}.
This analysis highlights the strengths and limitations of the algorithms in generating utility-preserving graphs under varying privacy requirements.

\textit{Triangle Counting.}
For Facebook and CA-HepPh, DP-dK exhibits significant fluctuations and higher relative error at lower privacy budgets, stabilizing only at $\varepsilon = 10$. 
In contrast, the others maintain consistently relative error across all privacy budgets.
For the ER Graph, TmF owns very low relative error across all privacy budgets, while DP-dK and DGG have higher errors, suggesting limitations in this specific context.

\textit{Degree Distribution.}
DP-dK consistently outperforms other methods across most of graphs, achieving the lowest KL divergence at higher $\varepsilon$ values. Other methods like PrivGraph and Tmf show varied performance, generally improving as $\varepsilon$ increases but not to the extent of DP-dK. 

\textit{Diameter.}
In general, DGG maintains a low and consistent RE across most of graphs and privacy budgets.
For all graphs, DP-dK have the highest relative errors than others in diameter. 
For the ER Graph, TmF, PrivSKG, and PrivHRG own the lowest relative error, which is equal to 0 approximately.

\textit{Community Detection.}
In most of graphs, DP-dK and PrivSKG achieve highest NMI values than others, which means that they can preserve the community structure very well.
In contrast, PrivHRG performs the worst for all graphs and privacy budgets.
The performance of TmF can be improved as the privacy budget increases, i.e., when $\varepsilon$ = 10.
Tmf and PrivSKG  maintain moderate MAE values but show improvement with higher $\varepsilon$ levels. 

\textit{Eigenvector Centrality.}
DP-dK demonstrates a steep decline in MAE with increasing $\varepsilon$, achieving the lowest errors across all datasets when $\varepsilon = 10$. 
Both PrivGraph and PrivSKG maintain moderate MAE values, but show improvement with higher $\varepsilon$ values. 
For the ER graph, the influence of privacy budgets on most algorithms, excluding PrivHRG, is minimal.

\textbf{Takeaways.} 
TmF consistently achieves high utility across various datasets and privacy budgets.
It reduces the added noise by the high-pass filtering technique.
DGG demonstrates particular strength in preserving utility in specific datasets such as Facebook and Wiki.
Its performance is comparable to TmF in several cases.
This is because the degree information is vital for most queries.
PrivGraph excels in multiple metrics, particularly in preserving community structures and eigenvector centrality.
It strikes a balance between perturbation noise and information loss by leveraging community information.
DP-dK exhibits higher error rates and lower NMI scores in several cases, especially at lower privacy budgets, indicating potential limitations in utility preservation under strict privacy constraints.
PrivHRG and PrivSKG show mixed performance, with higher error rates in several metrics, highlighting areas where further optimization and research could enhance their utility preservation capabilities.

\begin{table}[t]
\small
   \caption{Comparison of Time and Space Complexity.}
	\label{tab:time and space complexity}
	\centering
 	\setlength{\tabcolsep}{4mm}{
		\begin{tabular}{|c|c|c|}
			\hline
    Algorithms & Time Complexity & Space Complexity\\ \hline
    DP-dK     &  $O(n^2)$ &  $O(n^2)$   \\
    TmF       & $O(n^2)$ & $O(n^2)$ \\
    PrivSKG   &   $O(n^2m)$  &  $O(n^2)$      \\
    PrivHRG   & $O(n^2\log{n})$ & $O(m+n)$ \\
    PrivGraph & $O(n^2)$ & $O(m+n)$ \\
    DGG       &   $O(n^2)$  & $O(n^2)$  \\  \hline     
   \end{tabular}
    \begin{tablenotes}
     \item[1] $n$: number of nodes \quad $m$: number of edges
   \end{tablenotes}
   }
\end{table}

\subsection{Time and Space Analysis}
In this part, we compare the performance of algorithms theoretically and empirically, including time and space cost.

\noindent\textbf{Theoretical Analysis.}
Table \ref{tab:time and space complexity} summarizes the theoretical results of time complexity and space complexity.

\textit{Time Complexity.}
DP-dK, TmF, PrivGraph, and DGG all have a time complexity of $O(n^2)$, where $n$ is the number of nodes in the graph. 
This quadratic complexity suggests that these algorithms should handle moderate-sized graphs efficiently but may struggle with extremely large graphs.
PrivSKG has a higher time complexity of $O(n^2m)$, indicating potential inefficiency for very large graphs with many edges.
PrivHRG has a slightly higher time complexity of $O(n^2\log n)$, indicating that it may be less efficient for very large graphs due to the additional logarithmic factor.

\textit{Space Complexity.}
DP-dK, TmF, PrivSKG, and DGG have a space complexity of $O(n^2)$, indicating substantial memory requirements for large graphs.
PrivGraph and PrivHRG are more space-efficient with a complexity of $O(m+n)$, where $m$  is the number of edges, making them more suitable for sparse graphs.

\begin{remark}
   We represent graphs as an adjacency matrix in re-implementing algorithms for efficient queries.
   Thus, the time complexity and space complexity are $O(n^2)$ for most algorithms (e.g., DP-dK, TmF, PrivSKG, and DGG).
\end{remark}

\noindent\textbf{Empirical Analysis.}
Table \ref{tab:time cost} presents the empirical time cost (in seconds) for running each algorithm on various graph datasets.
DP-dK consistently shows the lowest time cost across most datasets, indicating its efficiency in practice.
TmF and DGG also demonstrate reasonable time costs, making them practical for larger datasets.
PrivSKG has significantly higher time costs, particularly on larger datasets like ca-HepPh and ER graph, suggesting scalability issues.
The main reason is that PrivSKG has to spend additional time to compute the smooth sensitivity.
PrivGraph shows moderate time costs, balancing efficiency and performance.

Table \ref{tab:memory cost} provides a comparison of empirical memory consumption (in megabytes) for the algorithms.
PrivGraph is the most memory-efficient, particularly on smaller datasets like Minnesota and Facebook.
TmF, PrivSKG, and DGG generally require moderate memory, making them suitable for memory-constrained environments.
DP-dK consumes more memory than others, especially on larger datasets, indicating potential challenges in memory-limited scenarios.

\begin{table}[t]
\small
   \caption{Comparison of Time Cost (seconds).}
	\label{tab:time cost}
	\centering
		\begin{tabular}{|c|r|r|r|r|r|}
			\hline
           &  \multicolumn{5}{c|}{Algorithms} \\ \cline{2-6}
    Graphs & DP-dK  & TmF & PrivSKG  & PrivGraph & DGG\\ \hline    
    Minnesota &  0.12 & 9.28   & 252.72 & 0.88 & 0.11\\
    Facebook  & 1.36  & 27.83  & 9230.63 & 3.37 & 0.65 \\
    Wiki-Vote & 1.97  & 77.56  & 21833.8 & 7.05 & 1.21 \\
    ca-HepPh  & 9.58  & 207.97 & 43452.83 & 16.97 & 2.00 \\
    poli-large & 8.75 & 317.35 & 6721.03 & 21.33 & 2.22 \\
    Gnutella  & 4.65  & 688.26 & 22630.92 & 46.29 & 4.24 \\
    ER graph  & 4.27  & 164.86 & 46995.37& 16.38 & 1.58 \\
    BA graph  & 8.01  & 137.83 & 9230.20 & 10.54 & 0.95 \\
     \hline     
   \end{tabular}
\end{table}

\begin{table}[t]
\small
   \caption{Comparison of Memory Consumption (Megatypes).}
	\label{tab:memory cost}
	\centering
   \resizebox{1.04\columnwidth}{!}{
		\begin{tabular}{|c|r|r|r|r|r|}
			\hline
           &  \multicolumn{5}{c|}{Algorithms} \\ \cline{2-6}
    Graphs & DP-dK  & TmF & PrivSKG  & PrivGraph & DGG\\ \hline    
    Minnesota & 108.26 & 53.28   & 75.15 & 22.93 & 111.00\\
    Facebook  & 129.27  & 124.50 & 117.46 & 79.85 & 303.28\\
    Wiki-Vote & 156.93 &  386.29  & 327.08 & 184.49 & 846.83\\
    ca-HepPh  & 6649.7  & 1100.20 & 1200.01 & 461.97 & 2291.97\\
    poli-large & 8861.51 & 1850.87 & 1167.29 & 711.27 & 3730.51\\
    Gnutella  & 7821.59  & 3927.03 & 4640.66 & 1508.71 & 7913.61\\
    ER graph  & 1783.50  & 763.02 & 1245.09 & 379.87 & 1624.07\\
    BA graph  & 5600.40  & 763.02 & 1174.19 & 308.90 & 1562.60\\
     \hline     
   \end{tabular}
   }
\end{table}

\section{Conclusions}
\label{sec:conclusion}

We addressed the challenge of comparable empirical studies on differentially private synthetic graph generation algorithms. 
Through a comprehensive literature study, we identified key elements of existing studies, including mechanisms, graph datasets, privacy requirements, and utility metrics, and formulated design principles to ensure comparability. 
Based on these principles, we introduced $\mathsf{PGB}$, a benchmark that meets all principles for fair comparison. 
We conducted the largest empirical study on differentially private synthetic graph algorithms to date, revealing valuable insights into the strengths and weaknesses of existing mechanisms. 
Our study highlights that while some algorithms perform well under certain conditions, no single solution is universally optimal. 

\section*{Acknowledgment}
The authors would like to thank the anonymous reviewers for their helpful comments. 
This work was supported in part by JSPS KAKENHI JP23K24851, JST PRESTO JPMJPR23P5, JST CREST JPMJCR21M2, National Key RD Program of China (2022YFB3103401, 2021YFB3101100), NSFC (62102352, 62472378, U23A20306), Zhejiang Province Pioneer Plan (2024C01074).
Jinfei Liu serves as the corresponding author. 
Shang Liu contributed to this work when he was a research assistant at the Institute of Science Tokyo.

\bibliographystyle{IEEEtran}
\bibliography{IEEEabrv,Reference}

\newpage
\begin{appendices}
\end{appendices}

\section{Verification}
To ensure the reliability and correctness of our re-implemented code, we conducted a comprehensive verification process. 
This section presents a comparison of our results with those reported in the original papers, utilizing identical or closely matched experimental settings.

In our benchmark experiments, we evaluate the following methods: DP-dK \cite{wang2013preserving}, TmF \cite{nguyen2015differentially}, PrivSKG \cite{mir2012differentially}, PrivHRG \cite{xiao2014differentially}, PrivGraph \cite{yuan2023privgraph}, and DGG \cite{qin2017generating}. We utilized the original source code for both PrivHRG and PrivGraph. 
Additionally, we implemented DGG within a central setting, which differs from the local setting described in the original paper; therefore, a direct comparison between our results and those in the original paper is not feasible.
Considering the straightforward nature of the DGG implementation, we will not delve into it further in this section. 
Instead, we will focus on verifying the re-implementations of DP-dK, TmF, and PrivSKG.

\textbf{DP-dK.}
Table \ref{tab:DP-dK verification} presents the results from the original paper \cite{wang2013preserving} alongside those from our re-implementation of DP-dK. 
Since most of the datasets used in the original paper are unavailable, we conducted our evaluation on the CA-GrQC dataset. 
Overall, we observe that most of our results are similar to those reported in the original study. 
Interestingly, our re-implementation even shows improved performance on certain metrics, such as the assortativity coefficient, average clustering coefficient, and modularity.
One notable difference lies in the diameter results, which can be attributed to variations in the construction methods. 
After obtaining the private dK distribution, we used the Havel-Hakimi algorithm to generate synthetic graphs, whereas the original paper did not specify the construction algorithm used.

\textbf{TmF.} 
Since the original paper \cite{nguyen2015differentially} provides limited results, we instead compare our re-implementation with those from PrivGraph \cite{yuan2023privgraph}. 
To conserve space, we use the Facebook dataset as an example for verification. 
Fig \ref{fig:Verification_tmf_Degree_Distribution} and Fig. \ref{fig:Verification_tmf_community} show results for degree distribution and community detection. 
In the original figures, TmF results are shown with a red line and red inverted triangles; in our figures, TmF results are represented by an orange line with stars.
For the degree distribution, the Y-axis range in the original figures is [0,20], and for community detection, it spans [0,0.5]. 
Our observations reveal that the maximum, minimum, and general trends across both sets of line plots are comparable. 
For example, in degree distribution, the Y-axis range is around [10,15] in both cases, and both show a declining trend.

\textbf{PrivSKG.} 
Given that most datasets from the original paper are either unavailable or have since been updated, we conducted our evaluation on the CA-GrQC dataset. 
Specifically, we computed the degree distribution and average clustering coefficient, then compared the plots from our re-implementation to those in the original paper, as illustrated in Fig. \ref{fig:Verification_PrivSKG_Degree_Distribution} and Fig. \ref{fig:Verification_PrivSKG_ACC}.
Since the original paper did not provide exact values in the experimental section, we rely on comparisons of the maximum, minimum, and overall trends in the line plots. 
For instance, in Fig. \ref{fig:Verification_PrivSKG_Degree_Distribution}, both our and the original results show a maximum count of about $10^3$, with the count approaching zero at a degree of approximately 100. 
Both plots align with the power-law distribution pattern.

\section{Overall Results on Graph Queries}

We also evaluate the results of different algorithms on each query for all privacy budgets and graph datasets, which helps researchers find that which algorithm performs best for a specific type of graph queries.
As presented in Table \ref{tab:overall result on query}, each entry in the table indicates the number of times an algorithm achieved the best performance out of 6 $\varepsilon$ values and 8 graph datasets for a given query (Definition~\ref{def:overall metric on query}). 
The highest frequency in each case is highlighted in gray.
We can conclude some key findings from the overall results.

\begin{definition}
\label{def:overall metric on query}
Let $A$ be target algorithm.
Let $G = \{G_1, G_2, \ldots, G_{m}\}$ be a set of $m$ graph datasets. 
Let $E = \{\varepsilon_1, \varepsilon_2, \ldots, \varepsilon_{n}\}$ be a set of $n$ privacy budgets. 
Let $Q = \{Q_1, Q_2, \ldots, Q_{q}\}$ be a set of $q$ queries. 
Let \( B_{i} \)  be the best performance indicator:
   \[
   B_{jk} = 
   \begin{cases} 
   1 & \text{if } A \text{ performs best on } Q_i \text{ for } G_j \text{ and } \varepsilon_k \\
   0 & \text{otherwise}
   \end{cases}
   \]
Finally, we have:
   \[
   C_A(Q_i) = \sum_{j=1}^{m}\sum_{k=1}^{n} B_{jk},
   \]
where \( C_A(Q_i)  \) is the count of how often algorithm \( A \) performs best across $m$ graph datasets and $n$ privacy budgets for $Q_i$.
\end{definition}

According to Table \ref{tab:overall result on query}, we have the following observations.
1) DP-dK performs better than other methods when querying the degree distribution and average clustering coefficient.
It calibrates the noise based on the smooth sensitivity, achieving the strict differential privacy guarantee with smaller magnitude noise.
2) TmF achieves the highest counts on calculating the number of nodes, number of edges, average degree, modularity, assortativity coefficient, and eigenvector centrality.
It is because it leverages the high-pass filtering technique to avoid the whole matrix manipulation.
3) PrivSKG outperforms other methods for triangle counts, diameter, and global clustering coefficient.
It computes an private of a given graph in the stochastic Kronecker graph (SKG) model, achieving good results.
4) PrivHRG yields better outcomes than other approaches when assessing the community detection.
It infers the network structure by using a statistical hierarchical random graph (HRG) model, which is good for preserving the community structure.
5) PrivGraph demonstrates superior results compared to other methods when querying the number of nodes and eigenvector centrality.
The reason is that it reduces the excessive noise by exploiting the community information.
6) DGG shows higher performance than other methods on queries for the number of nodes, degree variance, average of all shortest paths, and distance distribution.
This is because DGG generates synthetic graphs using degree information, which is essential for most graph queries.

\section{Results of DER}
In our benchmark, we do not directly compare DER results, as DER is commonly considered a baseline approach relative to other methods, such as TmF and PrivGraph. 
However, to demonstrate DER's performance, we include a comparison with TmF and PrivGraph. 
As illustrated in Fig. \ref{fig:DER}, DER generally exhibits lower performance than the other methods.

\section{Sensitivity and Mechanisms}
Sensitivity \cite{dwork2014algorithmic} captures the amount of necessary noise to ensure differential privacy (DP). 
Two common sensitivity definitions are global sensitivity \cite{dwork2014algorithmic} and smooth sensitivity~\cite{nissim2007smooth}.

\begin{definition}[Global Sensitivity \cite{dwork2014algorithmic}]
	\label{def:global sensitivity}
    For a query function $f$: $D \rightarrow \mathbb{R}$, the global sensitivity is defined by 
       \begin{center}
           $\triangle_{GS}=\mathop{max}\limits_{D\sim D^\prime}|f(D)-f(D^\prime)|$,
       \end{center}
       where $D$ and $D^\prime$ are neighboring databases that differ in a single user's data.
\end{definition}

\begin{definition}[Smooth Sensitivity \cite{nissim2007smooth}]
	\label{def:smooth_sensitivity}
    For a query function $f: D \rightarrow \mathbb{R}$, the $\beta$-smooth sensitivity at a database $D$ is defined by 
       \begin{center}
           $S^\beta_f(D) = \mathop{max}\limits_{D^\prime \sim D} \left( \triangle_f(D^\prime) \cdot e^{-\beta \cdot d(D,D^\prime)} \right)$,
       \end{center}
       where $\triangle_f(D^\prime)$ is the local sensitivity at $D^\prime$ given by $\triangle_f(D^\prime) = \mathop{max}\limits_{D^{\prime\prime} \sim D^\prime} |f(D^\prime) - f(D^{\prime\prime})|$, $D$ and $D^\prime$ are neighboring databases differing in a single user's data, and $d(D, D^\prime)$ is the distance between $D$ and $D^\prime$.
\end{definition}

The Laplace Mechanism \cite{dwork2006calibrating} satisfies the requirements of differential privacy (DP) by adding random Laplace noise to the aggregated results. 
The magnitude of the noise is determined by the sensitivity $\Delta f$, i.e., global sensitivity.
It is defined as the maximum change in the output of the aggregation function $f$ when the input data $D$ is modified.
When $f$ is a numeric query, the formal definition is as follows:

\begin{definition}[Laplace Mechanism]
	\label{def:laplace}
Given any function $f: D \rightarrow R^k$,
let $\Delta f$ be the sensitivity of function $f$.
$M(x)=f(x)+(Y_1,...,Y_k)$ satisfies $\varepsilon$-differential privacy, where $Y_i$ are $i.i.d$ random variables drawn from Lap($\Delta f/\varepsilon$) and $\varepsilon$ is the privacy budget.
\end{definition}

While the Laplace mechanism is effective for handling numeric queries, it is not suitable for queries with non-numeric values. 
Exponential mechanism \cite{mcsherry2007mechanism} is applied whether a function’s output is numerical or categorical.
The formal definition is described as follows:

\begin{definition}[Exponential Mechanism]
    Given any quality function $q: (D \times O) \rightarrow R$, and a privacy budget $\varepsilon$, the exponential mechanism $M(D)$  outputs $o \in O$ with probability proportional to $\mathsf{exp}(\frac{\varepsilon q(D,o)}{2\Delta q})$, where $\Delta q = \max\limits_{\forall o, D \simeq D'} |q(D, o) - q(D', o)|$ is the sensitivity of the quality function.
    $M(D)$ satisfies $\varepsilon$-differential privacy under the following equation.
    \begin{equation}
        \mathsf{Pr}[M(D)=o]=\frac{\mathsf{exp}(\frac{\varepsilon q(D,o)}{2\Delta q})}{\sum_{o^\prime\in O}\mathsf{exp}(\frac{\varepsilon q(D,o^\prime)}{2\Delta q})}
    \end{equation}
\end{definition}

\begin{table*}[t]
\small
	\caption{Verification of DP-dK on CA-GrQC.}
	\centering
	\label{tab:DP-dK verification}
	\begin{tabular}{|l|c|c|c|c|c|c|c|}
			\hline
      \multirow{3}{*}{Query} & \multirow{3}{*}{Ground Truth}& \multicolumn{6}{c|}{$\varepsilon$}\\
    \cline{3-8}
    & & \multicolumn{2}{c|}{20} & \multicolumn{2}{c|}{2} &  \multicolumn{2}{c|}{0.2} \\ \cline{3-8}
    & & Original & Our & Original & Our & Original & Our \\
\hline
   $|\text{V}|$ & 5241 & 5242 & 5242.1 & 5239 & 5269.6 & 5382 & 5802.3 \\ \hline
   $|\text{E}|$ & 14484 & 14509 & 14442 & 14596 & 15456 & 19430 & 24260 \\ \hline
   $\overline{d}$ & 5.527 & 5.535 & 5.52 & 5.572 & 5.802 & 7.220 & 9.617 \\ \hline
   Ass & 0.659 & -0.018 & 0.902 & -0.007 & 0.889 & -0.005 & 0.827 \\ \hline
   ACC & 0.529 & 0.007 & 0.566 & 0.008 & 0.597 & 0.015 & 0.563 \\ \hline
   $l_{max}$ & 17 & 13 & 893 & 12 & 583 & 10 & 723 \\ \hline
   $\triangle$ & 48260 & 628 & 40120 & 745 & 48758 & 3035 & 159457 \\ \hline
   Transitivity & 0.629 & 0.008 & 0.525 & 0.009 & 0.52 & 0.017 & 0.486 \\ \hline
   Mod & 0.801 & 0.404 & 0.902 & 0.402 & 0.889 & 0.323 & 0.826 \\ \hline
    
\end{tabular}
     \vspace{-10pt}
\end{table*}

\begin{table*}
   \caption{Overall Results on Graph Queries.}
	\label{tab:overall result on query}
	\centering
 	\setlength{\tabcolsep}{5mm}{
    \begin{threeparttable}
       \resizebox{2.1\columnwidth}{!}{
		\begin{tabular}{|c|c|c|c|c|c|c|c|c|c|c|c|c|c|c|c|}
			\hline
     Algorithms & \multicolumn{15}{c|}{Graph Queries} \\ \cline{2-16}
     & $|\text{V}|$  & $|\text{E}|$ & $\triangle$  & $\overline{d}$ & $d_{\sigma}$ & $\bm{d}$ & $l_{max}$ & $\overline{l}$ & $\bm{l}$ & GCC  & ACC & CD & Mod & Ass & EVC \\ \hline
     DP-dK & 0 & 0 & 6 & 0 & 9& \cellcolor{gray!50}35& 0 &0 &7 &12 & \cellcolor{gray!50}29 & 0& 14& 3 &3  \\ \hline
     TmF & \cellcolor{gray!50}48 & \cellcolor{gray!50}48 & 11 & \cellcolor{gray!50}48 & 6 & 4 & 10 & 12 & 10 & 10 & 9 & 11 & \cellcolor{gray!50}21 & \cellcolor{gray!50}16 & \cellcolor{gray!50}11 \\ \hline
     PrivSKG & 0	& 0	& \cellcolor{gray!50}20 & 0 &	8&	0&\cellcolor{gray!50}17&	7&	4 &\cellcolor{gray!50}19 &0	& 0&	0&	10&	7 \\ \hline
     PrivHRG & 6& 0&	2&	0&	2&	0&	14&	5&	8&	3&	4&	\cellcolor{gray!50}37&	5&	15&	9 \\ \hline
     PrivGraph & \cellcolor{gray!50}48&	0&	3&	0&	0&	0&	6&	5&	3&	4&	0&	0&	2&	3&	\cellcolor{gray!50}11 \\ \hline
     DGG &\cellcolor{gray!50}48&	0&	6&	0&	\cellcolor{gray!50}23&	9&	15&	\cellcolor{gray!50}19&	\cellcolor{gray!50}16&	0&	6&	0&	6&	1&	7\\ \hline 
   \end{tabular}
   }
    \begin{tablenotes}
    \item[1] Each number shows how often the algorithm performs best across 6 privacy budgets and 8 datasets. \\
    For example, the first number '48' in the second row means that TmF outperforms others in all cases.
    \item[2] The highest frequency in each case is highlighted in gray.
   \end{tablenotes}
   \end{threeparttable}
   }
     \vspace{-10pt}
\end{table*}

\begin{figure*}[t]
 \begin{center}
     \begin{minipage}[t]{0.49\linewidth}
   \centering
\subfigure[Original]{
		\begin{minipage}[t]{0.49\linewidth}
			\centering
			\includegraphics[width=\linewidth]{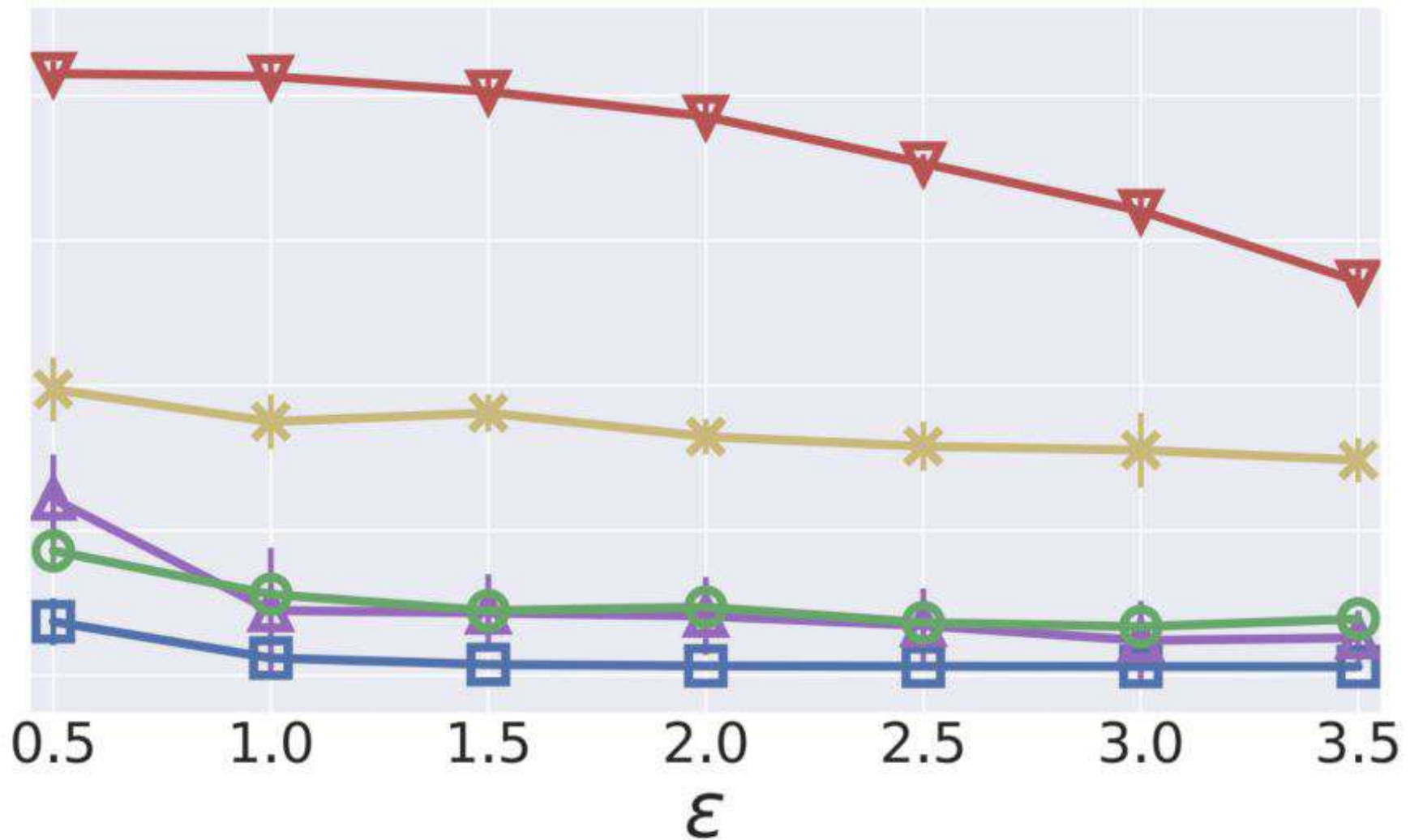}
		\end{minipage}
	}%
\subfigure[Our]{
		\begin{minipage}[t]{0.49\linewidth}
			\centering
			\includegraphics[width=\linewidth]{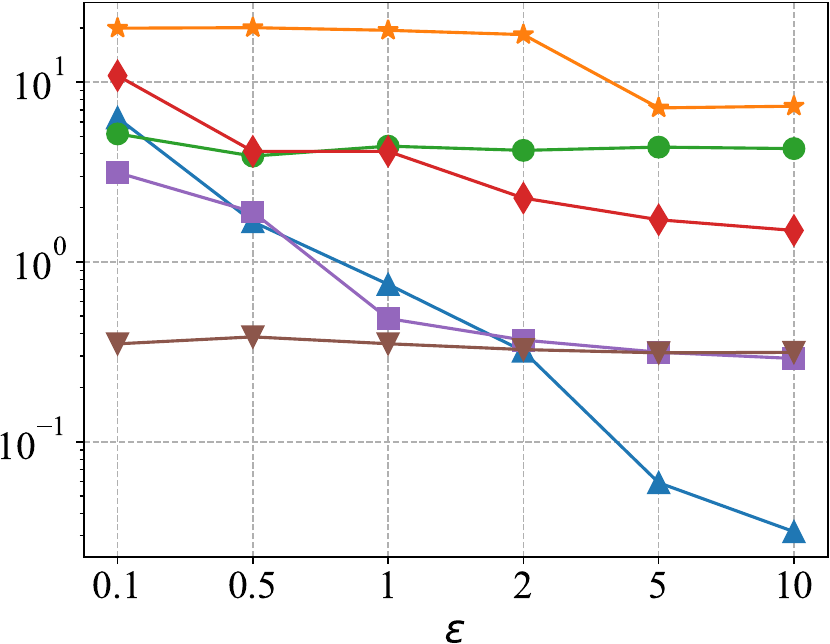}
		\end{minipage}%
	}
       \caption{Degree Distribution of TmF.}
       \label{fig:Verification_tmf_Degree_Distribution}
\end{minipage}
 \begin{minipage}[t]{0.49\linewidth}
   \centering
\subfigure[Original]{
		\begin{minipage}[t]{0.49\linewidth}
			\centering
			\includegraphics[width=\linewidth]{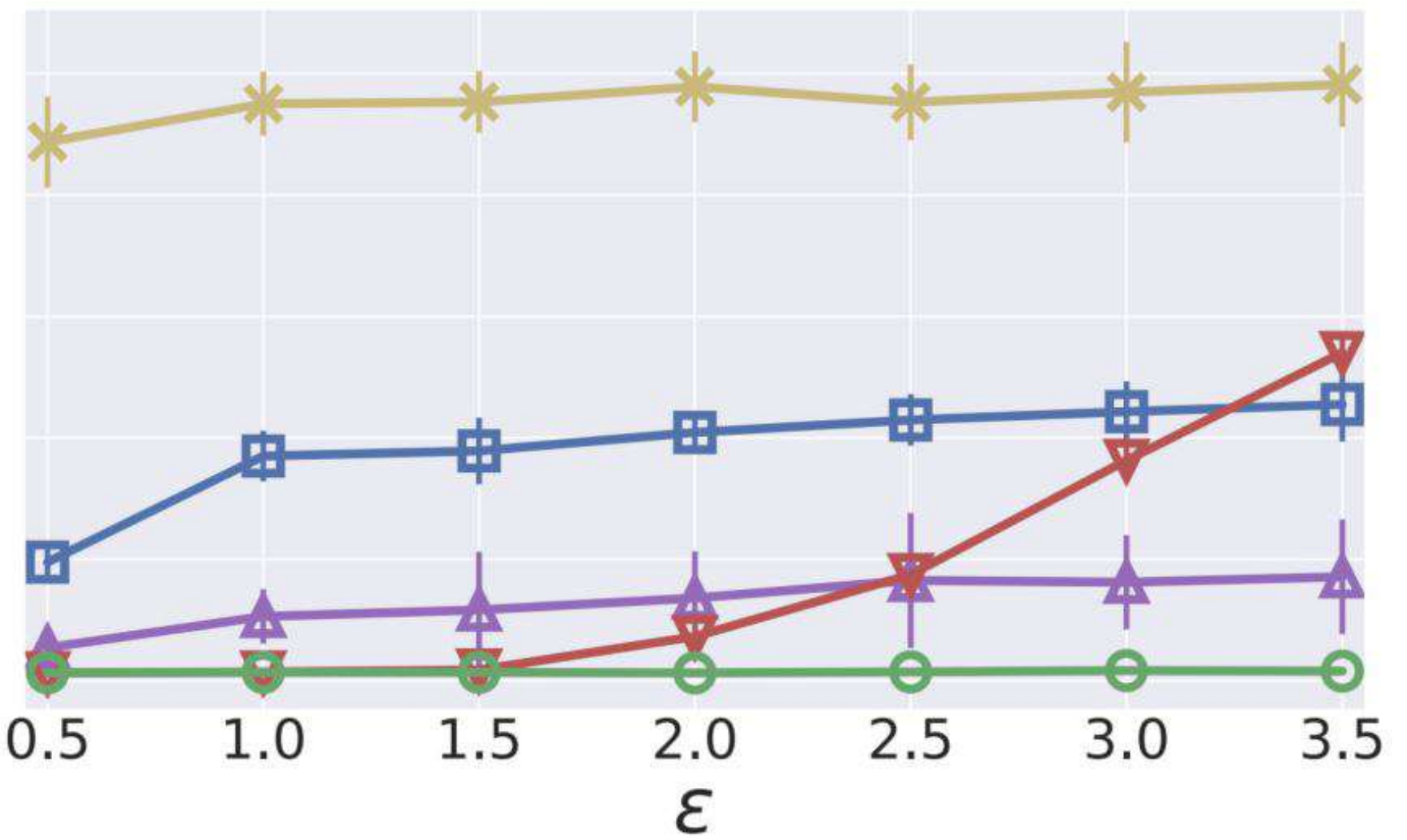}
		\end{minipage}
	}%
\subfigure[Our]{
		\begin{minipage}[t]{0.49\linewidth}
			\centering
			\includegraphics[width=\linewidth]{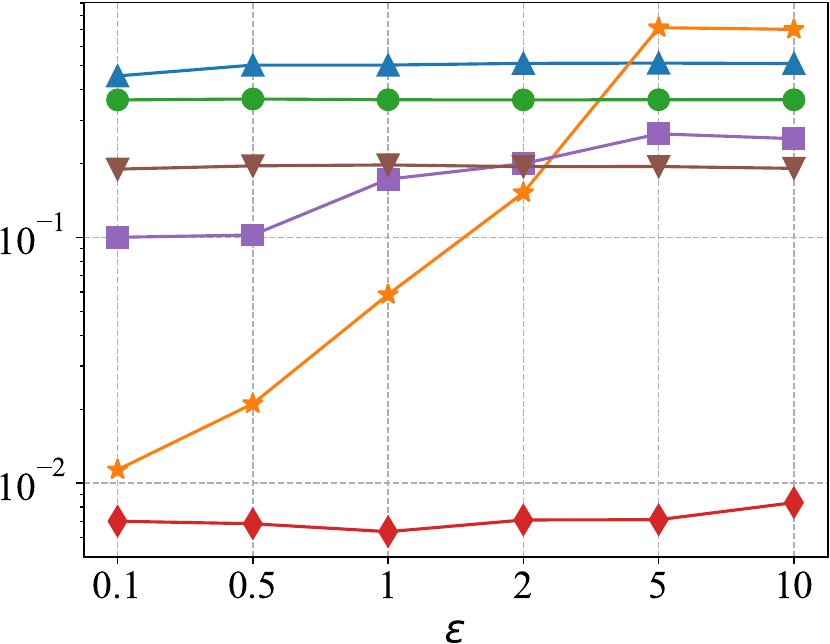}
		\end{minipage}%
	}
       \caption{Community detection of TmF.}
       \label{fig:Verification_tmf_community}
\end{minipage}
   \end{center}
      \vspace{-0.5cm}
\end{figure*}

\begin{figure*}[t]
 \begin{center}
     \begin{minipage}[t]{0.49\linewidth}
   \centering
\subfigure[Original]{
		\begin{minipage}[t]{0.49\linewidth}
			\centering
			\includegraphics[width=\linewidth]{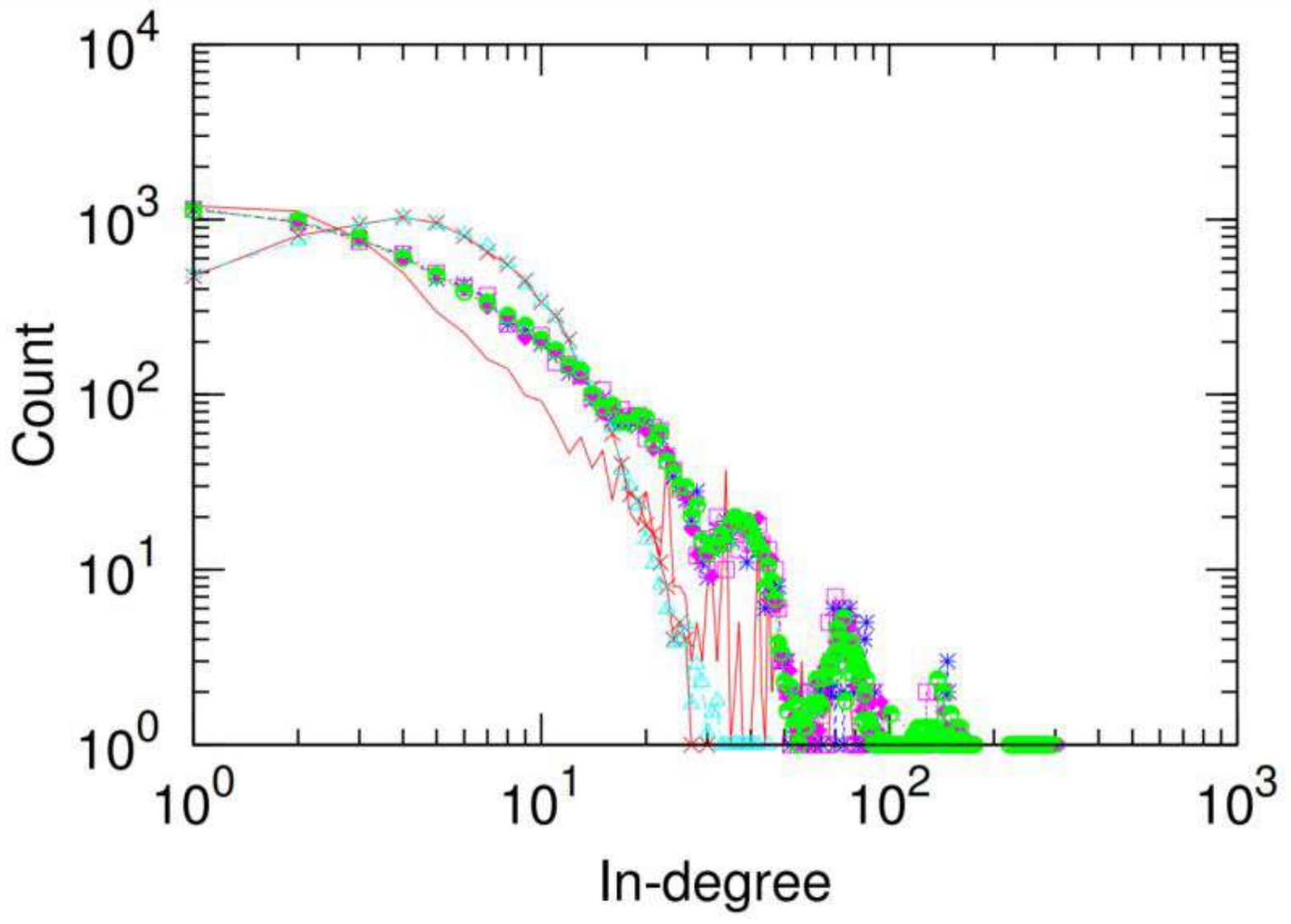}
		\end{minipage}
	}%
\subfigure[Our]{
		\begin{minipage}[t]{0.49\linewidth}
			\centering
			\includegraphics[width=\linewidth]{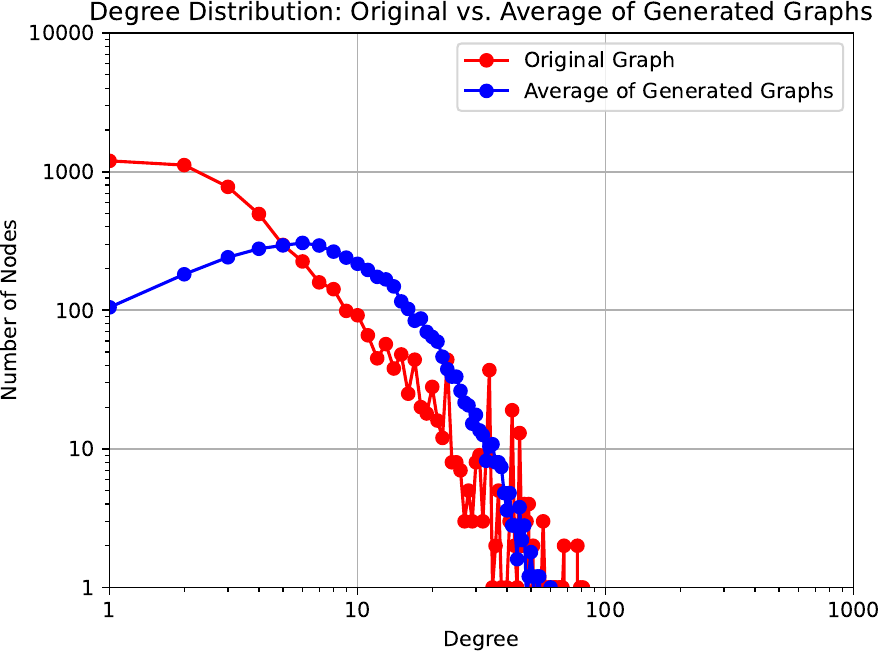}
		\end{minipage}%
	}
       \caption{Degree Distribution of PrivSKG.}
       \label{fig:Verification_PrivSKG_Degree_Distribution}
\end{minipage}
 \begin{minipage}[t]{0.49\linewidth}
   \centering
\subfigure[Original]{
		\begin{minipage}[t]{0.49\linewidth}
			\centering
			\includegraphics[width=\linewidth]{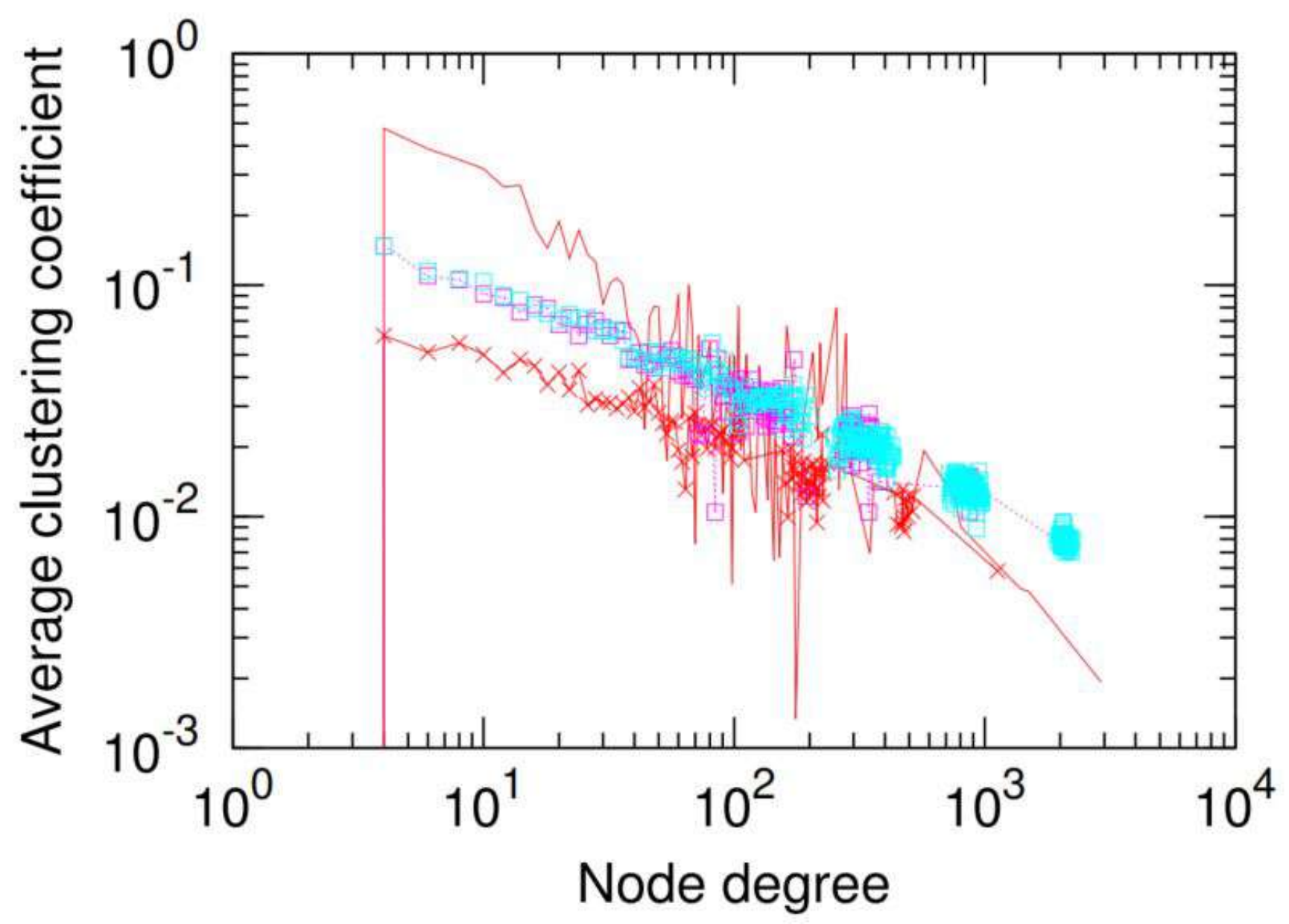}
		\end{minipage}
	}%
\subfigure[Our]{
		\begin{minipage}[t]{0.49\linewidth}
			\centering
			\includegraphics[width=\linewidth]{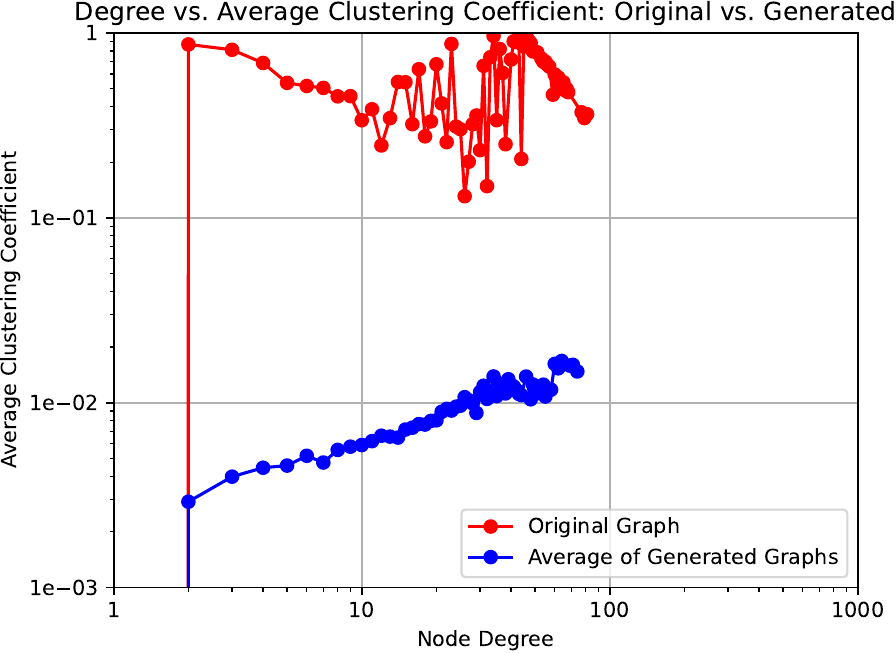}
		\end{minipage}%
	}
      \vspace{-0.4cm}
       \caption{Average Clustering Coefficient of PrivSKG.}
       \label{fig:Verification_PrivSKG_ACC}
\end{minipage}
   \end{center}
      \vspace{-0.5cm}
\end{figure*}

\begin{figure*}[h]
	\centering 
 	\subfigure{
		\begin{minipage}[t]{\linewidth}
			\centering
			\includegraphics[width=0.35\linewidth]{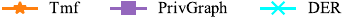}
               \vspace{-20pt}
		\end{minipage}
	}%
  \quad
	\subfigure{
		\begin{minipage}[t]{0.23\linewidth}
			\centering
			\includegraphics[width=\linewidth]{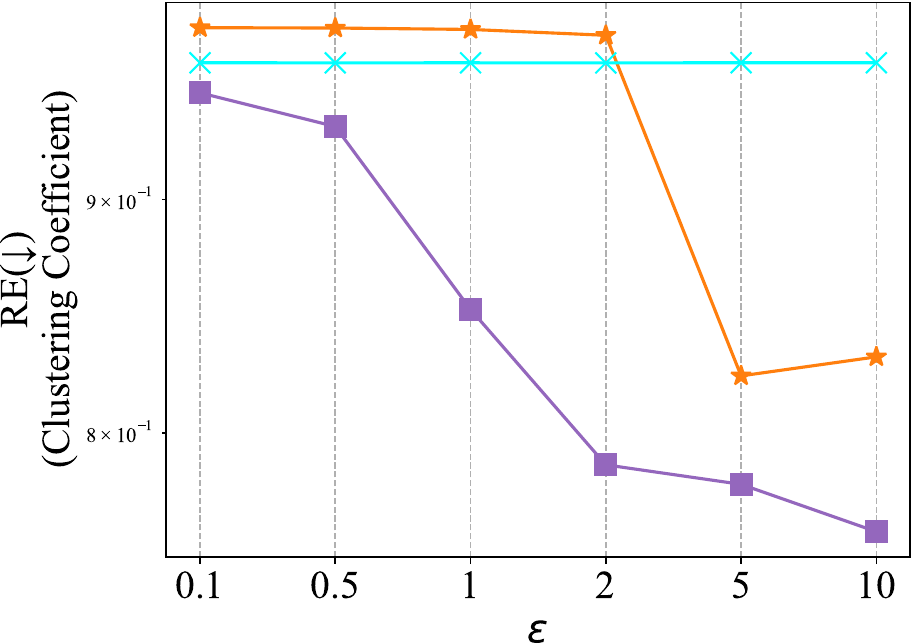}
            \vspace{-15pt}
		\end{minipage}
	}%
	\subfigure{
		\begin{minipage}[t]{0.23\linewidth}
			\centering
			\includegraphics[width=\linewidth]{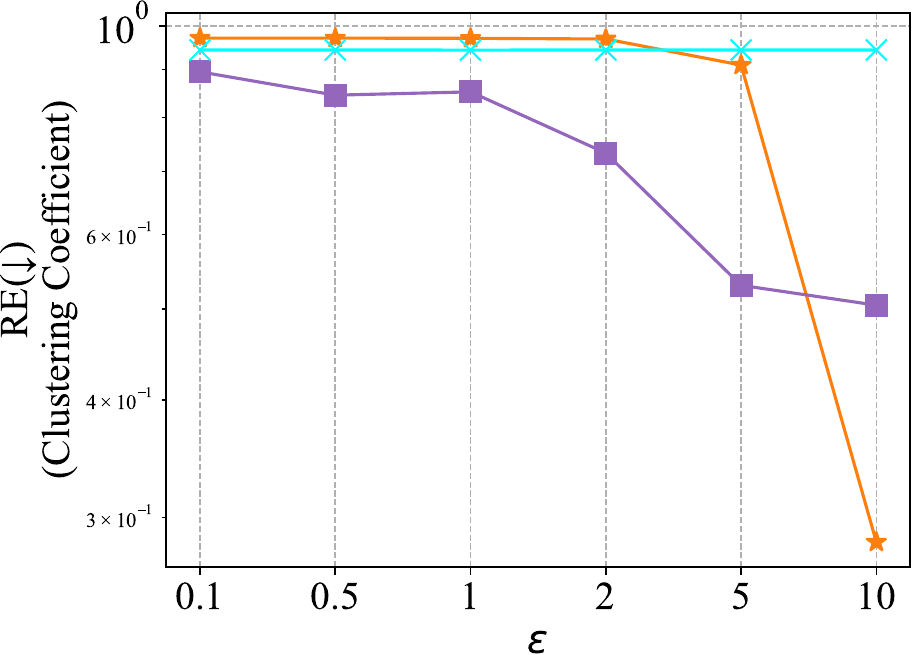}
               \vspace{-15pt}
		\end{minipage}%
	}
	\subfigure{
		\begin{minipage}[t]{0.23\linewidth}
			\centering
			\includegraphics[width=\linewidth]{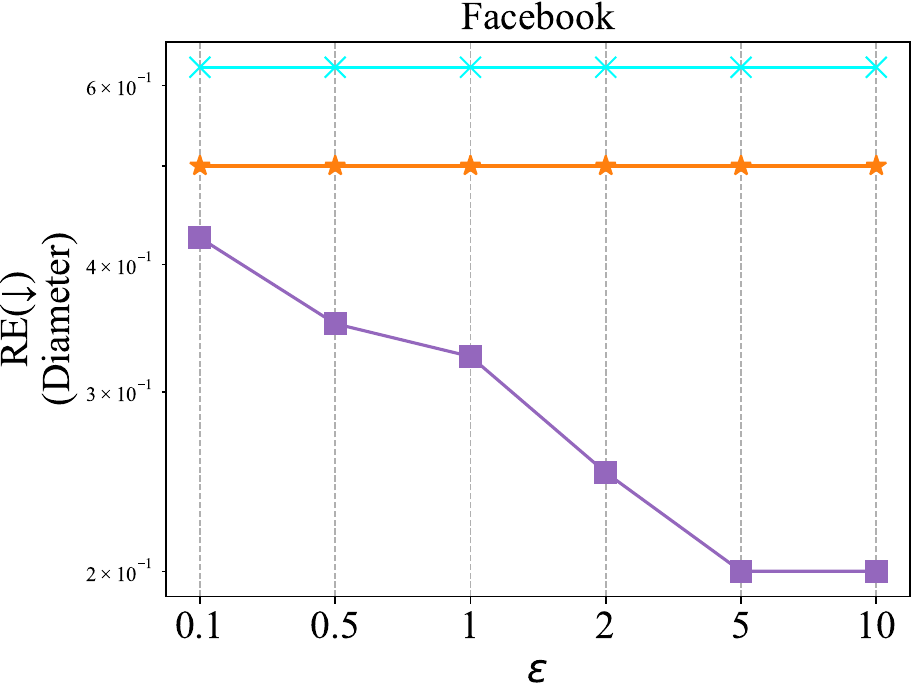}
               \vspace{-15pt}
		\end{minipage}%
	}
	\subfigure{
		\begin{minipage}[t]{0.23\linewidth}
			\centering
			\includegraphics[width=\linewidth]{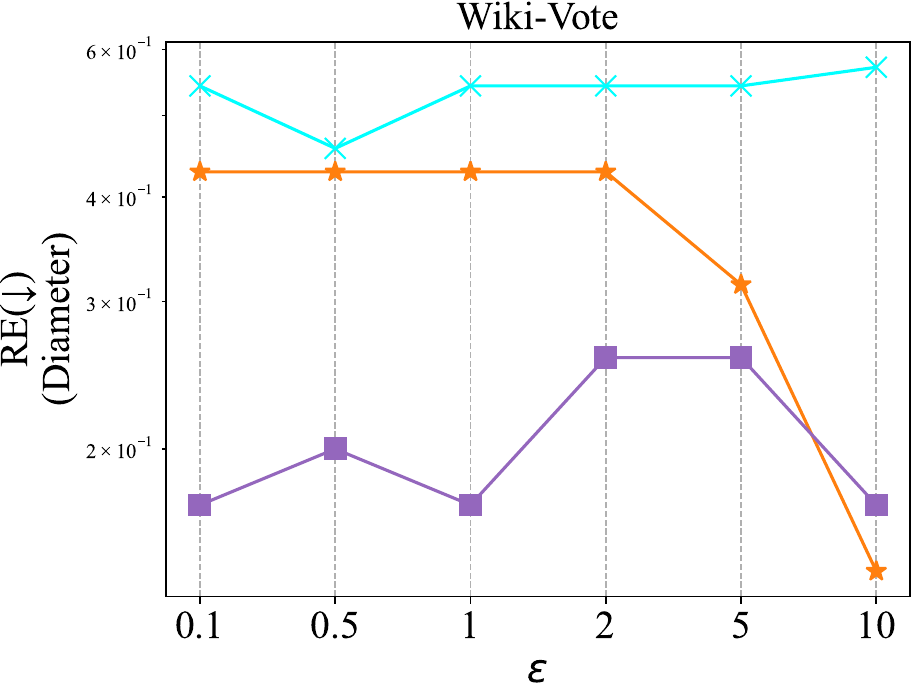}
               \vspace{-15pt}
		\end{minipage}%
	}
\centering
\vspace{-10pt}
\caption{End-to-end comparison of TmF, PrivGraph, and DER.
}
\label{fig:DER}
\vspace{-8pt}
 \end{figure*}


\end{document}